\documentclass[11pt,leqno,textwidth=10cm]{amsart}
\usepackage{amssymb}

\usepackage{mathrsfs}
\usepackage{mathtools}
\usepackage{abstract}

\usepackage{etoolbox}
\makeatletter
\patchcmd{\@maketitle}{\newpage}{}{}{} 
\makeatother

\usepackage{lmodern}            
\usepackage{float}							
\usepackage[pdftex]{graphics}    
\usepackage{graphicx}            
\usepackage{amsmath,amssymb,amsthm}  
\usepackage{tensor}             
\usepackage{braket}
\usepackage{dsfont}             
\usepackage[bf,format=plain]{caption}         
\usepackage{siunitx} 			  
\usepackage{color}
\usepackage{fancybox}
\usepackage[colorlinks,linkcolor=blue,citecolor=red,urlcolor=blue]{hyperref}    
\usepackage[a4paper,width=150mm,top=25mm,bottom=25mm,bindingoffset=6mm]{geometry}
\usepackage{caption}
\usepackage{subcaption}
\usepackage{makecell}
\usepackage{pdfpages}
\usepackage{nicefrac}

\newtheoremstyle{fancy}{}{}{\itshape}{}{\textsc\bgroup}{.\egroup}{ }{}
\newtheoremstyle{fancy2}{}{}{\rm}{}{\textsc\bgroup}{.\egroup}{ }{}
\theoremstyle{fancy}
\newcounter{intro}

\numberwithin{equation}{section}    

\newtheorem{lem}[equation]{Lemma}

\newtheorem{thm}[equation]{Theorem}

\newtheorem{named}[equation]{\name}
\newcommand{\name}{Proof of}

\theoremstyle{fancy2}
\newtheorem{dfn}[equation]{Definition}
\newcounter{axa}

\newtheorem{rem}[equation]{Remark}

\newcommand{\cref}[1]{Corollary~\ref{#1}}

\newcounter{subequation} 
 \newlength\mtabskip\mtabskip=-1.25cm
   
	 \def\mtabLong{long} 
 

\theoremstyle{plain}
\newtheorem{conjecture}{Conjecture}

\newcommand{\eq}[1]{\begin{equation}#1\end{equation}}

\newcommand{\bc}{\begin{cases}}
\newcommand{\ec}{\end{cases}}

\begin{document}


\title[Recollapsing spacetimes with $\Lambda<0$]{Recollapsing spacetimes with $\Lambda<0$}
\author[D.~Fajman, M.~Kraft]{David Fajman, Maximilian Kraft}


\date{\today}


\address{
David Fajman, Maximilian Kraft\newline
Faculty of Physics, 
University of Vienna\\ \newline
Boltzmanngasse 5\\ \newline
1090 Vienna, Austria\\ \newline
David.Fajman@univie.ac.at
}

\maketitle
\begin{abstract}
We show that any homogeneous initial data set with $\Lambda<0$ on a product 3-manifold of the orthogonal form $(F\times \mathbb S^1,a_0^2dz^2+b_0^2\sigma^2,c_0dz^2+d_0\sigma)$, where $(F,\sigma)$ is a closed 2-surface of constant curvature and $a_0,\hdots, d_0$ are suitable constants, recollapses under the Einstein-flow with a negative cosmological constant and forms crushing singularities at the big bang and the big crunch, respectively. Towards certain singularities among those the Kretschmann scalar remains bounded, hence these are not curvature singularities. We then show that the presence of a massless scalar field causes the Kretschmann scalar to blow-up towards both ends of spacetime for all solutions in the corresponding class. By standard arguments this recollapsing behaviour extends to an open neighborhood in the set of initial data sets and is in this sense generic close to the homogeneous regime. 
\end{abstract}


\section{Introduction}
The global-in-time behaviour of the Einstein flow remains one of the most intriguing open problems in Mathematical Relativity. While there has been significant progress in our understanding of the behaviour of solutions in the neighborhood of certain homogeneous attractors, both towards the complete expanding direction of spacetime as well as towards the collapsing direction, the global dynamics of the majority of cosmological spacetimes is still unknown.
The most general conjecture about a generic class of spacetimes is the \emph{Closed Universe Recollapse conjecture} stated by Barrow, Galloway and Tipler in 1986 \cite{barrow_galloway_tipler_conjecture}. 
 
\subsection{The closed universe recollapse conjecture}\label{chapter_closed_universe_recollapse}

A spacetime $(M,g)$ is said to \emph{recollapse} if all timelike geodesics in the maximal globally hyperbolic development are past and future incomplete \cite{ringstrom}.
As a consequence of Hawking's singularity theorem \cite{hawking_ellis_1973} a sufficient condition for recollapse of a spacetime is the existence of a foliation by CMC Cauchy hypersurfaces with crushing past and future singularities.
This means that the mean curvature $H$ of the leaves ranges from minus to plus infinity and thus has to vanish at some point. As $H$ is also a measure for the expansion rate of a spacetime, $H=0$ implies that the spacetime has reached a maximum of expansion. Given a spacetime foliated by CMC Cauchy hypersurfaces, a Cauchy hypersurface with vanishing mean curvature is called a \textit{maximal Cauchy hypersurface}.

\subsubsection{Origin of the conjecture}\label{historical_overview}
Originally, a conjecture was formulated by Zel'dovich and Grishchuk stating that universes admitting a compact Cauchy hypersurface without boundary recollapse and possess a maximal Cauchy hypersurface, cf. \cite{zeldovich1984}. The simplest example of a recollapse occurs for the Friedman universe with scalar curvature $\epsilon = 1$. It has spatial hypersurfaces topologically of the form $\mathbb S^3$.  Zel'dovich and Grishchuk gave an inhomogeneous example of a universe with spherical symmetry that recollapses. 
However, there are counterexamples to this conjecture, such as the Einstein-De Sitter solution on $\mathbb T^3$ \cite{rendall2008partial} or the Milne model on quotients of hyperbolic space \cite{AF20}. \\
More generally, it was shown by Barrow and Tipler that only spacetimes with topology $\mathbb S^3$ (or quotients thereof) and $\mathbb S^1 \times \mathbb S^2$ (or connected sums of those topologies) can admit a maximal Cauchy hypersurface, cf. \cite{Barrow1985}. They also showed if a globally hyperbolic spacetime has a maximal Cauchy hypersurface it is also future and past incomplete and the corresponding singularities are crushing, provided that $Ric(V,V)\geq 0$ for timelike vectors $V$ and a genericity condition holds. In a subsequent paper Barrow, Tipler and Galloway summarized their and other results and formulated three conjectures, cf. \cite{barrow_galloway_tipler_conjecture}, which are today referred to as \textit{closed universe recollapse conjecture}. 

\begin{conjecture}[\cite{barrow_galloway_tipler_conjecture}]\label{conjecture_1}
Every globally hyperbolic closed vacuum universe with either $\mathbb S^3$ or $\mathbb S^1 \times \mathbb S^2$ spatial topology begins in an initial singularity, expands to a maximal hypersurface and then recollapses to a final singularity.
\end{conjecture}
\begin{conjecture}[\cite{barrow_galloway_tipler_conjecture}]\label{conjecture_2}
Every globally hyperbolic spatial homogeneous closed matter-filled universe with either $\mathbb S^3$ or $\mathbb S^1 \times \mathbb S^2$ as spatial topology, satisfying the strong energy condition (\ref{strong_energy_condition}), the positive pressure-sum condition (\ref{non_negative_pressure_sum}), the dominant energy condition (\ref{dominant_energy_condition}) and the matter being regular (i.e. the evolution will not stop because of matter losing regularity, like shell-crossings), begins in an initial singularity, expands to a maximal hypersurface and then recollapses to a final singularity.
\end{conjecture}
\begin{conjecture}[\cite{barrow_galloway_tipler_conjecture}]\label{conjecture_3}
Every globally hyperbolic closed matter-filled universe with either $\mathbb S^3$ or $\mathbb S^1 \times \mathbb S^2$ spatial topology, satisfying the strong energy condition (\ref{strong_energy_condition}) and the positive pressure-sum condition (\ref{non_negative_pressure_sum}), begins in an initial singularity and ends in a final singularity. Hence, if all singularities are crushing or strong-curvature singularities, then the spacetime expands from the initial singularity to a maximal hypersurface and contracts to a final singularity again.
\end{conjecture}


It turns out that if matter is present the choice of the matter model is relevant. There are Friedman universes which do not recollapse, although having $\mathbb S^3$ topology, cf. \cite{barrow_galloway_tipler_conjecture}. In particular, one needs an additional assumption to the usual energy conditions (weak, strong, dominant). One has to assume at least that the sum of the principle pressures is non-negative.
However, the non-negative pressure-sum condition is necessary to ensure that the matter does not generate forces which could prevent a recollapse. If the pressures (or their sum) are negative, then the matter model intuitively contributes to the expansion of the spacetime rather than its contraction.


Tipler and Galloway proved Conjecture \ref{conjecture_2} for the FLRW spacetimes in \cite{barrow_galloway_tipler_conjecture} and it was shown by Gowdy that the Gowdy $\mathbb S^3$ and $\mathbb S^1 \times \mathbb S^2$ vacuum spacetimes recollapse, cf.~ \cite{Gowdy_2014}. 
Burnett later showed that any spherically symmetric spacetime possessing a $\mathbb S^1\times \mathbb S^2$ Cauchy hypersurface has timelike curves with finite lengths. It turned out to be easier to prove that timelike curves have finite lengths than showing the existence of a maximal hypersurface, cf.  \cite{Burnett1993}. This lead him to the following conjecture:
\begin{conjecture}[ \cite{Burnett1993}]\label{conjecture_4}
The lengths of timelike curves are finite for every globally hyperbolic closed matter-filled universe with either $\mathbb S^3$ or $\mathbb S^1 \times \mathbb S^2$ spatial topology, satisfying the dominant and non-negative pressure condition.
\end{conjecture}
This is a weaker version of the previous conjectures. It was further shown by Burnett that Conjecture \ref{conjecture_4} is true also for any spherically symmetric spacetime with $\mathbb S^3$ Cauchy hypersurface with dust as matter (Tolman spacetime), cf. \cite{Burnett1993}, and for the massless scalar field, cf. \cite{Burnett1994_2}. Finally, it could be shown by Burnett that Conjecture \ref{conjecture_4} holds for any spherically symmetric spacetime, cf. \cite{Burnett1994}.
Burnett and Rendall then proved that spherically symmetric spacetimes with a $\mathbb S^1 \times \mathbb S^2$ Cauchy surface, with Vlasov matter and the massless scalar field as matter models, have a constant mean curvature foliation and possess a maximal Cauchy hypersurface and hence proving the conjecture in this class, cf. \cite{burnett_1996} and \cite{Rendall_1995b}.
\begin{table}[ht]
    \centering
    \resizebox{\textwidth}{!}{\begin{tabular}{c|c|c|c|c}
    Spacetime & Matter & \makecell{Energy \\ Condition} & conjecture & Reference \\
    \hline
    \makecell{Gowdy \\ ($\mathbb S^1 \times \mathbb S^2$, $\mathbb S^3$)} & perfect fluid & \ref{non_negative_pressure_sum},\ref{dominant_energy_condition} & \ref{conjecture_1},\ref{conjecture_3}  & \cite{barrow_galloway_tipler_conjecture},\cite{Gowdy_2014}\\
    \hline
    \makecell{Kantowski-Sachs \\ ($\mathbb{R} \times \mathbb S^2$,$\mathbb S^1 \times \mathbb S^2$)} & \makecell{perfect fluid, \\ Vlasov} & \ref{non_negative_pressure}, \ref{strong_energy_condition}, \ref{dominant_energy_condition} & \ref{conjecture_2}  & \cite{Collins1977},\cite{Fajman_2019}\\
     \hline
     \makecell{Tolman\\ ($\mathbb S^3$)} & dust & \ref{non_negative_pressure}, \ref{dominant_energy_condition} & \ref{conjecture_2}& \cite{zeldovich1984} \\
    \hline
    FLRW ($\mathbb S^3$) & any & \ref{non_negative_pressure_sum}, \ref{dominant_energy_condition}, \ref{strong_energy_condition} & \ref{conjecture_1},\ref{conjecture_2}& \cite{barrow_galloway_tipler_conjecture} \\
    \hline
    Bianchi IX ($\mathbb S^3$) & any & \ref{non_negative_pressure_sum}, \ref{dominant_energy_condition} & \ref{conjecture_1},\ref{conjecture_2}& \cite{lin_wald_bianch_ix},\cite{heinzle_2010} \\
    \hline
    \makecell{spherically symmetric \\ ($\mathbb S^1 \times \mathbb S^2$, $\mathbb S^3$)} & any & \ref{non_negative_pressure}, \ref{dominant_energy_condition} & \ref{conjecture_4}& \cite{Burnett1994} \\
    \hline
   \makecell{spherically symmetric \\ ($\mathbb S^1 \times \mathbb S^2$)} & \makecell{Vlasov, MSF, \\ dust} & \ref{non_negative_pressure}, \ref{dominant_energy_condition} & 

   \ref{conjecture_1},\ref{conjecture_3}&
 \makecell{\cite{Rendall_1995b},\cite{burnett_1996} \\ \cite{rein_1996},\cite{Humphreys2012}} \\ 
    \hline
    \makecell{spherically \\ surface symmetric \\ ($\mathbb S^1 \times \mathbb S^2$)} & Vlasov, MSF & \ref{non_negative_pressure}, \ref{strong_energy_condition}, \ref{dominant_energy_condition} &  \ref{conjecture_1},\ref{conjecture_3} &\makecell{\cite{heinkeli}} \\   
    \end{tabular}}
    \caption{Known recollapsing spacetimes. MSF stands for massless scalar field.}
    \label{tab_closed_universe_recollapse}
\end{table}
Based on the results of Rendall and Burnett, it was shown by Henkel that a prescribed mean curvature foliation exists for spherically symmetric spacetimes with Vlasov matter, proving also conjecture \ref{conjecture_2}, cf. \cite{heinkeli} for this particular class of spacetimes. A similar result for massive Vlasov matter has been obtained by Rein, cf.~\cite{rein_1996}, also showing that for this class the singularities are curvature singularities.

Lin and Wald proved Conjecture \ref{conjecture_2} for general Bianchi IX spacetimes with matter satisfying the non-negative pressure-sum and the dominant energy condition, cf. \cite{lin_wald_bianch_ix}. In the vacuum case the result can be improved to an open subset of Bianchi type IX initial data \cite{ringstrom}. Additionally, it was shown by Calegero and Heinzle that the non-negative pressure-sum conditions is necessary for Bianchi IX to recollapse, as they were able to construct a family of spacetimes satisfying only the strong energy condition and do not recollapse, cf. \cite{heinzle_2010}. Calogero and Heinzle described Bianchi IX spacetimes with dynamical system analysis in great detail, cf. \cite{CalogeroHeinzele2011},  for massless (and massive) Vlasov matter, cf. \cite{Fajman_2019}.

Humphrey et al.~gave a complete overview over spherically symmetric dust-filled spacetimes, where they also showed Conjecture \ref{conjecture_2} for dust as matter (excluding solutions with shell-crossing and similar pathological behavior common for dust), cf.~\cite{Humphreys2012}.  The known recollapsing spacetimes are summarized in Table \ref{tab_closed_universe_recollapse}. We do not claim this list to be exhaustive.

Let us give some remarks on Table \ref{tab_closed_universe_recollapse}. The Kantowski-Sachs spacetimes cover both topologies $\mathbb{R} \times \mathbb S^2$ and $\mathbb S^1 \times \mathbb S^2$, cf. \cite{Collins1977} or \cite{Fajman_2019}. Nearly all results so far have been obtained for spatially homogeneous spacetimes. The only spacetimes appearing in Table \ref{tab_closed_universe_recollapse} which are not spatially homogeneous are the surface symmetric and spherically symmetric ones. 

Interestingly, for those spacetimes it is sufficient that the pressure normal to the surfaces of symmetry is positive, cf.~\cite{burnett_1991}, which is also the reason why the massless scalar field is a valid model. That the pressures are normal to the surfaces of symmetry means only the first eigenvalue of the stress tensor has to be positive. Hence, the non-negative pressure condition or pressure-sum condition can in principle be violated and a recollapse could still occur. An example of such a matter model is the massless scalar field, as it is shown in \cite{burnett_1996} and \cite{Rendall_1995b}. This raises the question if the assumptions on the matter can be relaxed for non-spatially homogeneous spacetimes with less symmetry.
Also in $2+1$ dimensions recollapse of spacetime occurs. In the corresponding models the decisive factor is the mass of particles in the relevant matter model. In particular, spacetimes with massive collisionless particles do not recollapse  \cite{F16} while those with massless collisionless particles recollapse, when the spatial topology is $\mathbb{S}^2$ \cite{Fajman_2017}. 

\subsection{The effect of a cosmological constant}\label{other_recollapses}
As the choice of the matter model affects the occurrence of a final singularity, it is natural to ask whether recollapse can be enforced with a suitable matter model. The most important assumption on matter for recollapses to appear is the non-negative pressure (sum) condition. The simplest example of a matter model with non-negative pressure is a negative cosmological constant. In such a case the Einstein equations take the form
\begin{align*}
    G_{\mu \nu} + \Lambda g_{\mu \nu} = 0 \quad \Leftrightarrow \quad G_{\mu \nu}  = - \Lambda g_{\mu \nu} \, .
\end{align*}
One may interpret the term $- \Lambda g_{\mu \nu}$ as a positive pressure term of a perfect fluid, when $\Lambda<0$. For the other energy conditions we observe the following. Without loss of generality we can choose a timelike vector $v$ such that $g(v,v) = -1$, then
\begin{align*}
    - \Lambda g(v,v) = \Lambda \le 0 \, ,
\end{align*}
which shows that the weak energy condition is not satisfied for $\Lambda<0$. The strong energy condition is satisfied, as
\begin{align*}
    \frac{1}{2} \si{tr}(-\Lambda g) g(v,v) = 2 \Lambda <  - \Lambda g(v,v) = \Lambda \, ,
\end{align*}
provided $\Lambda<0$. We consider spacetimes with negative cosmological constants in the remainder of this paper.
\subsubsection{Anti-deSitter spacetimes}
It is known that all globally hyperbolic vacuum Einsteinian spacetimes with a negative cosmological constant have a finite timelike diameter, meaning that all timelike geodesics are incomplete \cite{BEE96}. In this weak sense all spacetimes with $\Lambda<0$ recollapse. However, this does not provide detailed information on the global geometry of those spacetimes.

The simplest example of a cosmological spacetime with negative cosmological constant is the spatially closed Anti-de Sitter type space ($\Lambda=-3$), 
\eq{
\left((0,\pi)\times M,-dt^2+\sin(t)^2\cdot \gamma\right),
}
where $(M,\gamma)$ is a compact and closed Riemannian manifold without boundary of constant negative curvature $-1$. This spacetime indeed recollapses with crushing singularities, since the mean curvature is $H=-3\tan^{-1}(t)$. Maximally globally hyperbolic spacetimes of AdS type have been considered by Andersson, Barbot, Beguin and Zeghib, who have shown that a CMC time function exists taking on all real values, cf. \cite{Andersson2007}.

As remarked in \cite{Andersson2007}, the existence of CMC time function is stronger than the existence of a CMC foliation. In particular, a CMC time function implies a CMC foliation, which is the unique CMC foliation of the spacetime. Hence, this class of Anti-de Sitter spacetimes has crushing singularities and recollapses. 

The canonical solution with a positive cosmological constant is de Sitter spacetime, which is future and past complete and hence has no singularities. There are, however, spacetimes with a positive cosmological constant that do recollapse (cf.~\cite{Heinzle_2011}, \cite{Fajman_2016}). In \cite{Anderson_1997}, general $2+1$-dimensional spacetimes with $H^2$ as spatial topology where considered with an arbitrary cosmological constant. In the case of a positive cosmological constant the spacetimes have only an initial crushing singularity, while in the case of a negative cosmological constant the spacetimes have an initial and a final crushing singularity. $2+1$-dimensional spacetimes with $H^2$ spatial topology and positive cosmological constant expand forever for any initial data \cite{Fajman_2020} .

While positivity of the cosmological constant allows for all global scenarios (completeness and recollapse), in the case of a negative cosmological constant, to our present knowledge, all known examples are recollapsing with crushing initial and final singularities. Hence, it may be tempting to investigate further classes of spacetimes with a negative cosmological constant with respect to their global geometry to strengthen investigate the generality of this behaviour.  

\subsection{Main result: Recollapsing product spacetimes with $\Lambda<0$}
In the present paper we consider spacetimes with spatial product topology of the form $F\times \mathbb S^1$ with orthogonal, homogeneous initial data with a negative cosmological constant $\Lambda<0$. These homogeneous spacetimes allow for a rich dynamical behaviour. As our main result, we show that all spacetimes of this form recollapse under the Einstein flow with $\Lambda<0$ and give a classification of the state space of the corresponding dynamical system including the type of initial and final singularities of all solutions. The main theorem reads as follows.

\begin{thm}
All spatially homogeneous anisotropic vacuum or scalar field spacetimes with orthogonal initial data of the form $(F\times \mathbb S^1,a_0^2dz^2+b_0^2\sigma,c_0dz^2+d_0\sigma)$ and $\Lambda < 0$ possessing a CMC hypersurface of topology $\mathbb S^1 \times F$, with either $F = \mathbb S^2$, $F=H^2$ or $F=\mathbb S^1 \times \mathbb S^1$ (or compact quotients thereof) admit a CMC foliation. The mean curvature of the leaves takes on all real values, producing future and past crushing singularities. In the presence of a nontrivial scalar field, all singularities are curvature singularities.
\end{thm}
\begin{rem}
We conclude by a standard argument based on the local-stability of the Einstein equations and Hakwing's singularity theorem analogously to \cite{ringstrom} that initial data sets sufficiently close to the homogeneous data considered above all generate recollapsing spacetimes.
\end{rem}

\subsection{Overview on the paper}
We introduce the geometric setup in Section \ref{sec:prel} and discuss the case of the vacuum Einstein flow in Section \ref{section_vacuum}. In Section \ref{section_curvature_singularity} we present the case of a non-vanishing massless scalar field and in Section \ref{section_remarks} we deduce the behaviour of initial data close to the homogeneous case. The appendix contains various standard lemmas on dynamical systems we use throughout the paper. 
\subsection*{Acknowledgements}
D.F.~acknowledges the support of the Austrian Science Fund (FWF) through the  Project \emph{Relativistic fluids in cosmology} (P 34313-N). The authors thank Greg Galloway and Liam Urban for comments. The graphs where made using the open source software \textit{Maxima}.

\section{Preliminaries}\label{sec:prel}

In this chapter we derive the effective equation for the vacuum Einstein flow with negative cosmological constant for spacetimes of the form $\mathbb S^1 \times F$, with $F$ being a surface of constant scalar curvature $\epsilon$ and metric of the form $-dt^2+a(t)^2dz^2+b(t)^2\sigma_\varepsilon$.

\subsection{Field equations}
We assume that $\Lambda < 0$, hence $|\Lambda|= - \Lambda$. Thus, the Einstein vacuum equations with cosmological constant read
\begin{align}
    R_{\mu \nu} - \frac{R}{2} g_{\mu \nu} = |\Lambda| g_{\mu \nu} \, .
\end{align}
For the metric we choose the following ansatz using Gau\ss\,coordinates, i.e. setting lapse to one and shift to zero, which yields
 \begin{align*}
     g = - \si{d}t^2 + a^2 \si{d}z^2 + b^2 \sigma_\epsilon \, ,
 \end{align*}
 with the scale factors $a=a(t),b=b(t)$, which encode the dynamical behaviour of spacetime. $(F,\sigma_\epsilon)$ is a two dimensional Riemannian manifold with constant scalar curvature $\epsilon$. The scalar curvature is normalised to the cases $\epsilon=-1,0,1$. 

 In the case $\epsilon = -1$ we have that $F \simeq H^2$, which is the hyperbolic plane or constant quotients thereof. Such spacetimes are known as Bianchi Type III spacetimes. If $\epsilon = 1$ we have that $F \simeq \mathbb S^2$, which is the $2$-dimensional sphere. In this case the spacetime falls into the class of Kantowski-Sachs spacetimes. And for $\epsilon = 0$ we have $F \simeq T^2$, which is the flat $2$-torus and such spacetimes fall in the class of Bianchi type I.
 
 We proceed with calculating the evolution equations. The projection of the metric onto the spatial slices $F\times\mathbb S^1$ reads
  \begin{align*}
     h(t) = a(t)^2 \si{d}r^2 + b(t)^2 \sigma_\epsilon \, .
 \end{align*}
The Hamiltonian constraint and the two ADM evolution equations read
 \begin{align*}
    0 &= R_h - k_{ij}k^{ij} + \si{tr}k^2 + 2|\Lambda| \, , \\
    \partial_t h_{ij} &= -2 k_{ij} \, , \\
    \partial_t k_{ij} &= |\Lambda| h_{ij} + \left({\si{Ric}_h}\right)_{ij} + k_{ij} \, \si{tr}_h k - 2 k_{il} k\indices{^{l}_j} \, ,
 \end{align*}
 with $k$ being the second fundamental form
 \begin{align*}
     k= - \left( a \dot{a} \, \si{d}^2r + b \dot{b} \, \sigma_\epsilon \right) \, .
 \end{align*}
Furthermore the following equations hold.
 \begin{align*}
     R_h &=  \frac{2 \epsilon}{b^2} \, ,\quad &k_{ij}k^{ij} =& \frac{\dot{a}^2}{a^2} + 2 \frac{\dot{b}^2}{b^2} \, ,\\
      k_{1l} k\indices{^{l}_1} &= \dot{a}^2 \, , \quad &k_{2l} k\indices{^{l}_2} =& \dot{b}^2 \, , \\
      \si{tr}_h k &= - \left( \frac{\dot{a}}{a} + 2 \frac{\dot{b}}{b} \right) \, , \quad & \partial_t k_{11} =& - a \ddot{a} - \dot{a}^2 \, ,\\
       \partial_t k_{22}  &= - b \ddot{b} - \dot{b}^2 \, , \quad & \left({\si{Ric}_h}\right)_{22} =& - \epsilon \, ,
 \end{align*}
Inserting into the reduced system gives second order equations for $a$ and $b$, given by
  \begin{align}
      \Ddot{b} =& - \frac{\dot{b}^2}{b} - |\Lambda| b - \frac{\epsilon}{b} - \frac{\dot{a} \dot{b}}{a} \, , \label{evolution_b} \\
       \Ddot{a} =& - |\Lambda| a - 2 \frac{\dot{a} \dot{b}}{b} \, ,  \label{evolution_a_2} \\
       0 =& |\Lambda| + \frac{\epsilon}{b^2} + \frac{ \dot{b}^2}{b^2} + 2 \frac{\dot{b} \dot{a}}{b a} \label{constraint_ab} \, .
  \end{align}
  Combining the constraint with b gives a second order equation for $b$ only, 
  \begin{align}
      \Ddot{b} = - \frac{1}{2} \dot{b}^2 - \frac{1}{2} b |\Lambda| - \frac{\epsilon}{2} \frac{1}{b} \, . \label{evolution_onlyb}
  \end{align}
The Kretschmann scalar reads,
  \begin{align}
    R^{\alpha \beta \gamma \sigma}  R_{\alpha \beta \gamma \sigma} = \frac{8 {{{\Ddot{b}} }^{2}}}{{{b}^{2}}}+\frac{4 {{ {\dot{b}} }^{4}}}{{{b}^{4}}}+\frac{8 {{{\dot{a}} }^{2}} {{ {\dot{b}} }^{2}}}{{{a}^{2}} {{b}^{2}}}+\frac{8 {{ {\dot{b}} }^{2}}}{{{b}^{4}}}+\frac{4}{{{b}^{4}}}+\frac{4 {{{\Ddot{a}} }^{2}}}{{{a}^{2}}} \, . \label{Kretschmann_scalar}
\end{align}

\section{The vacuum Einstein flow with $\Lambda<0$} \label{section_vacuum}

\subsection{An explicit solution}
In the case $\epsilon = -1$ there exists an explicit solution of the form
\begin{align}
    a = \cos \left(\sqrt{|\Lambda|} t \right) \, , \qquad b = \frac{1}{\sqrt{|\Lambda|}} \, . \label{pancake_solution}
\end{align}
It can easily be verified, as
\begin{align*}
    \dot{a} = \sqrt{|\Lambda|} \sin \left(\sqrt{|\Lambda|} t \right) \, , \qquad \Ddot{a} = -|\Lambda| \cos \left(\sqrt{|\Lambda|} t \right) = -|\Lambda| a \, ,
\end{align*}
and the derivatives of $b$ vanish. Hence all equations are satisfied. The spacetime is expanding and then contracting into a pancake-like singularity -- one dimension remains finite while the two others collapse. Taking a look at the Kretschmann scalar (\ref{Kretschmann_scalar}) we find that it stays finite, but the mean curvature 
\[ \si{tr}_h k = - \left( \frac{\dot{a}}{a} + 2 \frac{\dot{b}}{b} \right) =  - \sqrt{|\Lambda|}  \tan \left(\sqrt{|\Lambda|} t \right)\] diverges for $t \rightarrow \pm \frac{\pi}{2} \sqrt{|\Lambda|}^{-1}$. Thus, the spacetime described by the solution recollapses and generates crushing singularities, which are however not curvature singularities.

\subsection{General homogeneous solutions}
The aim of the section is to prove the theorem:
\begin{thm}
All spatial homogeneous spacetimes of the form
\eq{
-dt^2+a(t)^2dr^2+b(t)^2\sigma_\varepsilon,
} with $\Lambda < 0$ possessing a CMC hypersurface of topology $\mathbb S^1 \times F$, with either $F = \mathbb S^2$, $F=H^2$ or $F=\mathbb S^1 \times \mathbb S^1$ or compact quotients thereof admit a CMC foliation. The mean curvature of the leaves takes on all real values, yielding future and past crushing singularities.
\end{thm}
\begin{proof}
The case $F = \mathbb S^2$ is shown by Lemma \ref{epsilon_1_recollapse}, $F = \mathbb S^1 \times \mathbb S^1$ is shown by Lemma \ref{epsilon_0_recollapse} and the $F = H^2$ is shown by Lemma \ref{epsilon_-1_recollapse}.
\end{proof}
What remains is showing the auxiliary lemmas. Before we proceed we reduce the Einstein equations into a first order system. We introduce the variables
\begin{align}
    H:= \frac{\dot{a}}{a} + 2 \frac{\dot{b}}{b}\, , \qquad \sigma := \frac{\dot{a}}{a} - 2 \frac{\dot{b}}{b} \, ,
\end{align}
where $H$ is in fact the mean curvature. It measures the overall expansion of the spacetime and $\sigma$ measures the shear. 

\subsubsection{Curvature $\varepsilon =-1$}
We start with the case $\epsilon=-1$. Thus, the evolution equations take the form
\begin{align}
    \dot{H} &= -1 - \frac{1}{4} H \sigma - \frac{3}{8} \left( H^2 + \sigma^2\right) \, , \label{evolution_H} \\
    \dot{\sigma} &= -1 - \frac{3}{4} H \sigma - \frac{5}{8} H^2 + \frac{3}{8} \sigma^2 \, . \label{evolution_sigma}
\end{align}
And equation (\ref{constraint_ab}) gives
\begin{align}
    1 + \frac{5}{16} H^2 - \frac{3}{16} \sigma^2 - \frac{1}{8} H \sigma =  e^{-2y} > 0 \, , \label{constrain_inequality}
\end{align}
which is an additional constraint to the state space of $H,\sigma$.

We now compactify the state space to use the theory of dynamical system analysis as summarized in Section \ref{subsection_cumpactification}. We start with introducing expansion-normalised dynamical variables as described in Section \ref{subsection_dynamical_systems}. The asymptotic regions will therefore turn into equilibrium points. Firstly, we define the expansion normalised time $\tau_{N}$ by demanding
\begin{align*}
    \frac{\si{d} t}{ \si{d} \tau_{N} } = \frac{1}{N} \, , \quad N:= \sqrt{1+\frac{1}{9} H^2} \, .
\end{align*}
Then we define new expansion normalised variables
\begin{align*}
   H_{N}:= \frac{1}{3} \frac{H}{N} \, , \quad \sigma_{N}:= \frac{1}{3} \frac{\sigma}{N} \, .
\end{align*}
One can easily check that $H \rightarrow \infty$ corresponds to $H_N = 1$ and $H \rightarrow -\infty$ corresponds to $H_N = -1$ by calculating the respective limits. Hence, the reparametrization compactifies state space in the sense that $H_N \in [-1,1]$. Though, $\sigma$ is not normalized to the interval $[-1,1]$. There may exist a better way to compactify the state space but as we are mainly interested in the behavior of $H$ (and $H_N$) it is sufficient for the following.
We define \[ S:= \left \{ (H_N, \sigma_N) \in \mathbb{R}^2 \; | \; -1 \le H_N \le 1\right \} \] as state space.

From the definitions follow the identities
\begin{align*}
    H = \frac{3 H_{N} }{\sqrt{1- H_N^2}} \, , \quad \frac{1}{1+\frac{1}{9} H^2} = 1- H_{N}^2 \, .
\end{align*}
Hence, using the previous identities we obtain
\begin{align*}
     \frac{\si{d} H_{N}}{ \si{d} \tau_{N} } &= \frac{\si{d} t}{ \si{d} \tau_{N} }  \frac{\si{d} H_{N}}{ \si{d} t } = \frac{1}{N}\frac{\si{d}}{ \si{d} t } \left( \frac{\frac{1}{3}H}{N}\right) = \frac{\frac{1}{3}\dot{H}}{1+\frac{1}{9} H^2} - \frac{1}{9} H \frac{\frac{1}{3}H \dot{H}}{(1+\frac{1}{9} H^2)^{2}} \\
     &= \frac{\frac{1}{3}\dot{H}}{1+\frac{1}{9} H^2} \left( 1 -  H_{N}^2 \right) 
      = \frac{\frac{1}{3}\dot{H}}{\sqrt{1+\frac{1}{9} H^2}} \frac{ 1 -  H_{N}^2}{\sqrt{1+\frac{1}{9} H^2}} = \frac{\frac{1}{3}\dot{H}}{\sqrt{1+\frac{1}{9} H^2}} \left( 1 -  H_{N}^2 \right)^{3/2}  \\
 \end{align*}   
and
 \begin{align*}     
      \frac{\si{d} \sigma_{N}}{ \si{d} \tau_{N} } &= \frac{\si{d} t}{ \si{d} \tau_{N} }  \frac{\si{d} \sigma_{N}}{ \si{d} t }  = \frac{1}{N}\frac{\si{d}}{ \si{d} t } \left( \frac{\frac{1}{3}\sigma}{N}\right) = \frac{\frac{1}{3}\dot{\sigma}}{1+\frac{1}{9} H^2} - \frac{1}{9} \sigma \frac{H \frac{1}{3}\dot{H}}{(1+\frac{1}{9} H^2)^{2}} 
     \\
         &= \sqrt{1-{{H_{N}}^{2}}} \frac{\frac{1}{3}\dot{\sigma}}{\sqrt{1+\frac{1}{9} H^2}} - H_{N} \sigma_{N}  \sqrt{1-{{H_{N}}^{2}}} \frac{\frac{1}{3}\dot{H}}{\sqrt{1+\frac{1}{9} H^2}} \, .
\end{align*}
Now we can use the evolution equations for $H$ and $\sigma$, (\ref{evolution_H}) and (\ref{evolution_sigma}) respectively, and insert them. After some calculation  one arrives at
\begin{align}
     \frac{\si{d} H_{N}}{ \si{d} \tau_{N} } = \left( 1-{{H_{N}}^{2}}\right)\left( -\frac{9}{8} \left( \sigma_{N}^2+H_{N}^2\right)-\frac{3}{4} H_{N} \sigma_{N}-\frac{1}{3} \left( 1-{H_{N}}^{2} \right) \right) \, , \label{expansion_normalised_H}
\end{align}
as well as
\begin{align}
\begin{split}
        \frac{\si{d} \sigma_{N}}{ \si{d} \tau_{N} }  =& \frac{9}{8} {{\sigma_{N}}^{2}}-\frac{9}{4} H_{N} \sigma_{N}-\frac{1}{3}\left(1-{{H_{N}}^{2}}\right)-\frac{15}{8} {H_{N}}^{2}  \\
    &- \sigma_N H_N \left( -\frac{9}{8} \left( \sigma_{N}^2+H_{N}^2\right)-\frac{3}{4} H_{N} \sigma_{N}-\frac{1}{3} \left( 1-{H_{N}}^{2} \right) \right)  \, . \label{expansion_normalised_sigma}
    \end{split}
\end{align}   
And the constraint inequality becomes
\begin{align}
    0 < \frac{1}{9} \left(1 - H_{N}^2 \right) + \frac{5}{16} H_{N}^2 - \frac{3}{16} \sigma_{N}^2 - \frac{1}{8} H_{N} \sigma_{N} \, . \label{expansion_normalised_constraint}
\end{align}
We introduce a more compact notation by calling the right-hand side of equation (\ref{expansion_normalised_H}) $F_1$ and the right-hand side of equation (\ref{expansion_normalised_sigma}) $F_2$. We obtain
\begin{align*}
     \frac{\si{d}}{ \si{d} \tau_{N} }  \begin{pmatrix}
 H_{N} \\
 \sigma_{N}  \\
\end{pmatrix} = 
\begin{pmatrix}
 F_1(H_{N}, \sigma_{N}) \\
 F_2(H_{N}, \sigma_{N})  \\
\end{pmatrix} \, .
\end{align*}
We search for equilibrium points by setting the derivatives to zero and solving the resulting equations for $H_{N}$ and $\sigma_{N}$, in particular by solving
\[\begin{pmatrix}
 F_1(H_{N}, \sigma_{N}) \\
 F_2(H_{N}, \sigma_{N})  \\
\end{pmatrix} =  0 \, .\]
We get six real equilibrium points, namely
\begin{align*}
    A &= \left(-1,\frac{5}{3} \right) \, , &\quad B &= \left(-1,1 \right) \, , &\quad C &= \left(-1,-1 \right) \, , \\
    D &= \left(1, -\frac{5}{3} \right) \, , &\quad E &= \left(1, -1 \right) \, , &\quad F &= \left(1,1 \right) \, .
\end{align*}
We will linearize the system of differential equations around the equilibrium points to obtain the qualitative behavior of the solutions. This is done by computing the matrix
\begin{align*}
   L = \begin{pmatrix}
 \frac{\partial F_1}{\partial H_{N}} & \frac{\partial F_1}{\partial \sigma_{N}} \\
 \frac{\partial F_2}{\partial H_{N}} & \frac{\partial F_2}{\partial \sigma_{N}} \\
\end{pmatrix}  
\end{align*}
of partial derivatives and evaluating it at the equilibrium points to obtain the linearized system
\begin{align*}
     \frac{\si{d}}{ \si{d} \tau_{N} }  \begin{pmatrix}
 H_{N} \\
 \sigma_{N}  \\
\end{pmatrix} =  L(x_1,x_2)
\left(\begin{pmatrix}
 H_{N} \\
 \sigma_{N}  \\
\end{pmatrix} -
\begin{pmatrix}
 x_1 \\
 x_2  \\
\end{pmatrix}
\right) \, .
\end{align*}
This gives an approximation of the solution around the equilibrium point $(x_1,x_2)$. We then calculate the eigenvalues and eigenvectors of the respective matrices at the equilibrium points.

We denote the eigenvalues by $\lambda_1$ and $\lambda_2$ for the first and second eigenvalue respectively. They are given in Table \ref{tab_fixed_points_1}. We see that the points $A$ and $C$ are sinks (past attractors), as their eigenvalues all have a negative real part. Additionally they have a stable $2$-dimensional manifold. The points $F$ and $D$ are sources (future attractors), as their eigenvalues have all a positive real part. Their unstable manifold is $2$-dimensional. The points $B$ and $E$ are saddle points, as they have eigenvalues with mixed sign. They have one stable $1$-dimensional manifold and one unstable $1$-dimensional manifold. The saddle point $B$ is attractive in direction of $H_{N}$ as the first eigenvalue is positive and repulsive in direction of $\sigma_{N}$ as the second eigenvalue is negative. The saddle point $E$ is attractive in direction of $\sigma_{N}$ and repulsive in direction of $H_{N}$. 
\begin{table}[ht]
    \centering
    \begin{tabular}{lr}
    
    \begin{tabular}{c|c|c}
    Fixed point & $\lambda_1$ & $\lambda_2$ \\
    \hline
    $A$ & -6 & -2 \\
    \hline
    $B$ & $\nicefrac{3}{2}$ & -3 \\
    \hline
    $C$ & -6 & -6 \\
    \end{tabular}
    &\quad
    \begin{tabular}{c|c|c}
    Fixed point & $\lambda_1$ & $\lambda_2$  \\
    \hline
    $D$ & 2 &6  \\
    \hline
    $E$ & $-\nicefrac{3}{2}$ &3 \\
    \hline
    $F$ & 6 &6  \\
    \end{tabular}
    \end{tabular}
    \caption{Eigenvalues of the fixed points $A,B,C,D,E,F$.}
    \label{tab_fixed_points_1}
\end{table}

Furthermore, the system has no periodic orbits, due to the following consideration: The evolution equation (\ref{expansion_normalised_H}) implies
\[  \frac{\si{d} H_{N}}{ \si{d} \tau_{N} } < 0 \, ,\] 
as long as $H_N < 1$. Hence, $H_{N}$ is forced to be monotonically decreasing. In particular, the monotonicity property of $H$ is inherited by $H_{N}$. But the existence of a periodic orbit would imply that $H_{N}$ changes its monotonicity. Hence, there cannot be any periodic orbits. 

The line given by $H_N-\sigma_N=0$ is a heteroclinic orbit (an orbit which connects two equilibrium points). It corresponds to the isocline $H=\sigma$ in the non-compactified state space. One can see this by applying Theorem \ref{theorem_invariant_sets}. Define $Z_1:=H_N - \sigma_N$. Then calculating the derivative of $Z_1$ with respect to $\tau_N$ and using (\ref{expansion_normalised_H}) and (\ref{expansion_normalised_sigma}) for derivatives of $H_N$, $\sigma_N$ respectively yields
\begin{align*}
\frac{\si{d}}{\si{d} \tau} Z_1  &= \alpha_1(H_N,\sigma_N) Z_1 \, ,
\end{align*}
with
\begin{align*}
     \alpha_1(H_N,\sigma_N):= \frac{1}{24} \left( 19 \, H_N^3 + 18 \, H_N \sigma_N + 27 \, H_N \sigma_N^2 + 26 \, H_N + 54 \, \sigma_N \right) 
\end{align*}
being a ${C}^1$-function in $\sigma$ and $H_N$ and also $\tau_N$. Hence the conditions of the theorem are satisfied and thus $\{Z_1 = 0\}$ is an invariant set. It defines a straight line ($\sigma_N=H_N$) in the state space which we call 
\begin{align*}
    O_1(H_N):=H_N \, .
\end{align*}
Also $Z_1>0$ and $Z_1<0$ are invariant sets, which means that the state space is divided into two regions, one for which $H_N > \sigma_N$ and one for which $H_N < \sigma_N$.

The saddle points $B$ and $E$ have one positive and one negative eigenvalue. Thus they have a $1$-dimensional stable manifold and a $1$-dimensional unstable manifold. Those correspond precisely to the only flow lines which can flow into (or out of) the saddle points. As the saddle points are separated by the line $O_1(H_N)$, the flow lines flowing into the saddle points can only emanate from $F$ or $C$ respectively. Such manifolds are invariant sets and to find them we use Theorem \ref{theorem_invariant_sets}.  Define
\begin{align}
    Z_2 :=  \frac{1}{27} \left( 19 H_N^2 + 8 \right)- \sigma_N^2\, . 
\end{align}
The derivative of $Z_2$ can be calculated by inserting equation (\ref{expansion_normalised_H}) and (\ref{expansion_normalised_sigma}) for the derivatives of $H_N$ and $\sigma_N$ respectively. It gives
\begin{align*}
    \frac{\si{d} Z_2 }{\si{d} \tau_N}     &= \alpha_2(H_N,\sigma_N) Z_2 \, ,
\end{align*}
with $\alpha_2$ being a ${C}^1$-function in $H_N$, $\sigma_N$ and $\tau_N$ given by
\begin{align*}
     \alpha_2(H_N,\sigma_N):= \frac{19}{12} H_N^3 + \frac{3}{2} H_N^2 \sigma_N + \frac{9}{4} H_N \sigma_N^2 - \frac{19}{12} H_N + \frac{9}{4} \sigma_N \, .
\end{align*}
Hence, applying Theorem \ref{theorem_invariant_sets} gives that $Z_2 = 0$ is an invariant set. We can solve the equation and obtain two curves. The curve
\begin{align}
    O_2(H_N):= \frac{1}{3 \sqrt{3}} \sqrt{19 H_N^2 + 8} \, , \label{hyperbolic_curves_1}
\end{align}
is connecting the saddle point $B$ with the source $F$ and the other one,
\begin{align}
    O_3(H_N):= -\frac{1}{3 \sqrt{3}} \sqrt{19 H_N^2 + 8} \, , \label{hyperbolic_curves_2}
\end{align}
connecting the sink $C$ with the saddle point $E$. The sets $Z_2 > 0$ and $Z_2 < 0$ are also invariant due to Theorem \ref{theorem_invariant_sets}. 

Solving the constraint inequality (\ref{expansion_normalised_constraint}) yields two hyperbolic curves which enclose state space. These are given by
\begin{align}
    \sigma_N = \pm \frac{\left(4 \; \sqrt{2 \; H_N^{2} + 1} + \sqrt{3} \; H_N \right)}{3 \; \sqrt{3}} \label{constraint_1} \, ,
\end{align}
and define two curves in the state space, namely
\begin{align*}
    O_4(H_N)&:=\frac{\left(4 \; \sqrt{2 \; H_N^{2} + 1} - \sqrt{3} \; H_N \right)}{3 \; \sqrt{3}} \, , \\
    O_5(H_N)&:=\frac{-\left(4 \; \sqrt{2 \; H_N^{2} + 1} + \sqrt{3} \; H_N \right)}{3 \; \sqrt{3}} \, .
\end{align*}
One can easily check that $O_4(1) = 1$ and $O_4(-1) = \frac{5}{3}$. Hence, $O_4$ starts in $F$ and ends in $A$. For the other curves we get that $O_5(-1) = -1$ and $O_5(1) = -\frac{5}{3}$. Hence, $O_5$ starts in $D$ and ends in $C$. 

Collecting the previous results shows that the state space is divided into four invariant sets, which we will call $ABF$, $BCF$, $CEF$ and $CED$. They are given by
\begin{align*}
    ABF:= \left\{ (H_N, \sigma_N) \in S \; | \; O_2(H_N) < \sigma_N <  O_4(H_N) \right\} \, , \\
    BCF:= \left\{ (H_N, \sigma_N) \in S \; | \; O_1(H_N) < \sigma_N <  O_2(H_N) \right\} \, ,  \\
    CEF:= \left\{ (H_N, \sigma_N) \in S \; | \; O_3(H_N) < \sigma_N <  O_1(H_N) \right\} \, , \\
    CED:= \left\{ (H_N, \sigma_N) \in S \; | \; O_5(H_N) < \sigma_N <  O_3(H_N) \right\} \, .
\end{align*}
From Theorem \ref{theorem_poincare} (Generalised Poincare-Bendixson) follows that all $\alpha$- or $\omega$-limit sets can only be either periodic orbits, equilibrium points or heteroclinic cycles. The system has no periodic orbits because $H_N$ has fixed monotonicity. Additionally, it cannot have heteroclinic cycles, as the existence of a heteroclinic cycle would require a sequence of saddle points connected with each other. But the saddle points $B=(-1,1)$ and $E=(1,-1)$ are separated from each other, as $B$ lies in the invariant set $H_N < \sigma_N$, while $E$ lies in the invariant set $H_N > \sigma_N$. We conclude that the system can only have equilibrium points as $\alpha$- or $\omega$-limit sets.

A qualitative picture of the compactified state space is shown in Figure \ref{figure_phasespace_bianch_iii_compact}. The equilibrium points are marked as black dots. The qualitative behavior of the flow is illustrated by the orange curves. The red curves denote the constraint curves given by $O_4$ and $O_5$.

We can now also determine the asymptotic behavior of the solutions for the six different equilibrium points $A$ to $F$. We start with the equilibrium points $C=(-1,-1)$ and $F=(1,1)$. For those we have that
\begin{align*}
    H_N = 1 = \sigma_N \, , \quad H_N = -1 = \sigma_N
\end{align*}
Hence, inserting the definitions gives
\begin{align*}
    \frac{H}{N} = \frac{\sigma}{N} \quad \Rightarrow \quad H = \sigma \, .
\end{align*}
As $H=\sigma$, it follows that $\dot{b} = 0$, hence $b$ is constant. Solving the differential equation for $H$ gives
\begin{align*}
    H = \frac{\dot{a}}{a} = - \tan(C+ t) \quad \Rightarrow \quad a = \cos \left(C+t \right) \, .
\end{align*}
This corresponds to the explicit pancake solution given in (\ref{pancake_solution}).

For the saddle points $B = (-1,1)$ and $E=(1,-1)$ we have that
\begin{align*}
     \sigma_N &= - H_N \Rightarrow  H = - \sigma\\
  \Rightarrow \quad   \frac{\dot{a}}{a} - 2 \frac{\dot{b}}{b} &= - \frac{\dot{a}}{a} - 2 \frac{\dot{b}}{b} \Rightarrow  \frac{\dot{a}}{a} = 0 \, .
\end{align*}
Hence $a$ is constant. From (\ref{evolution_H}) we get
\begin{align*}
    \dot{H} = - 1 + \frac{1}{4} H^2 - \frac{6}{8} H^2 = -1 -\frac{1}{2} H^2 \, ,
\end{align*}
and solving the ODE yields
\[ H = \sqrt{2} \tan \left( \frac{1}{2} \sqrt{2} (c_1 - t) \right) \, ,\]
with $c_1$ denoting the integration constant. The function $b$ is just given by \[\frac{\dot{b}}{b} = H \, ,\] which gives
\begin{align}
     b = c_2 \left(\operatorname{cos} \left( c_1 - t \right) \right)^{\sqrt{2}} \label{assymptotic_saddle_solution}\, ,
\end{align}
where $c_2$ denotes a collection of constants. Solutions approaching the saddle points will asymptotically behave like (\ref{assymptotic_saddle_solution}). 

For the remaining equilibrium points $A=(-1,\frac{5}{3})$ and $D=(1,-\frac{5}{3})$ we have the relations
\begin{align*}
    H_N =& -\frac{3}{5} \sigma_N\Rightarrow- \frac{3}{5} \sigma = H \, ,\\
    - \frac{3}{5} \frac{\dot{a}}{a} + \frac{6}{5} \frac{\dot{b}}{b} =&  \frac{\dot{a}}{a} + 2 \frac{\dot{b}}{b} \Rightarrow- 2 \frac{\dot{a}}{a} = \frac{\dot{b}}{b} \, .
\end{align*}
Inserting into the equation (\ref{evolution_H}) gives
\begin{align*}
    \dot{H} = -1 + \frac{1}{4} \, \frac{5}{3} H^2 - \frac{3}{8} H^2 - \frac{3}{8} \, \left(\frac{5}{3}\right)^2 H^2 = -1 -H^2 \, ,
\end{align*}
and solving it yields
\begin{align*}
    H = -\tan \left(c_1 + t\right) \, .
\end{align*}
As $- 2 \frac{\dot{a}}{a} = \frac{\dot{b}}{b}$ we have
\[ \frac{\dot{a}}{a} = - H = \tan \left(c_1 + t\right) \, ,\]
while $b$ is given by the relation
\[ - 2 \frac{\dot{a}}{a} = \frac{\dot{b}}{b} \, .\]
Solving it gives
\begin{align}
 a =\cos^{-1}\left( c_1 + t\right) \, , 
 \quad b =  C \cos^2\left(c_1 + t \right)  \, ,\label{assymptotic_others}
\end{align}
where $C$ denotes a numerical constant. We can see that solutions asymptotic to (\ref{assymptotic_others}) have one scale factor converging to zero while the other diverges. 

It remains to analyze what type of singularity occurs at $T_\pm$. If $b$ vanishes, the Kretschmann scalar, as given in (\ref{Kretschmann_scalar}), diverges, due to the term $\frac{1}{b^2}$. If only $a$ vanishes, it is not clear if the diverging terms cancel out. However, the mean curvature, which was denoted by $H$, always diverges. Hence we can conclude that all solutions to the system (\ref{evolution_b})-(\ref{constraint_ab}) recollapse and the singularities are at least crushing singularities.

By collecting all the previous results we have proven the first auxiliary lemma:

\begin{lem}\label{epsilon_-1_recollapse}
All solutions to the system (\ref{evolution_b}),(\ref{evolution_a_2}),(\ref{constraint_ab}) recollapse and their mean curvature takes on all real values, hence producing crushing singularities. The asymptotic behavior of the spacetimes can be grouped into five classes:
\begin{enumerate}
\item Solutions whose initial data lies in the set $BCF$ or $CEF$ are future and past asymptotic to (\ref{pancake_solution}). Such solutions have no curvature singularities as the Kretschmann scalar remains finite.
\item Solutions whose initial data lies in the set $ABF$ are past asymptotic to (\ref{pancake_solution}) and future asymptotic to (\ref{assymptotic_others}). The future singularity is of Kasner type in the sense that one scale factor vanishes while the other diverges. Additionally, the future singularity is also a curvature singularity, as the Kretschmann scalar diverges.
\item Solutions whose initial data lies in the set $CED$ are past asymptotic to (\ref{assymptotic_others}) and future asymptotic to (\ref{pancake_solution}). The past singularity is of Kasner type and the Kretschmann scalar diverges, giving rise to a curvature singularity.
\item There exists one solution whose initial data is determined by (\ref{hyperbolic_curves_1}), which is past asymptotic to (\ref{pancake_solution}) and future asymptotic to (\ref{assymptotic_saddle_solution}). The future singularity is also a curvature singularity.
\item There exists one solution whose initial data is determined by (\ref{hyperbolic_curves_2}), which is past asymptotic to (\ref{assymptotic_saddle_solution}) and future asymptotic to (\ref{pancake_solution}). The past singularity is also a curvature singularity. 
\end{enumerate}
\end{lem}
\subsubsection{Curvature $\varepsilon=0$}
For the case $\epsilon = 0$ (Euclidean symmetry) we obtain from the system (\ref{evolution_b}),(\ref{evolution_a_2}) and (\ref{constraint_ab}) the equations
\begin{align}
      \Ddot{b} =& - \frac{\dot{b}^2}{b} - |\Lambda| b  - \frac{\dot{a} \dot{b}}{a} \, , \label{epsilon_0_equation_1}\\
       \Ddot{a} =& - |\Lambda| a - 2 \frac{\dot{a} \dot{b}}{b} \, ,  \label{epsilon_0_equation_2} \\
       0 =&\, |\Lambda| + \frac{ \dot{b}^2}{b^2} + 2 \frac{\dot{b} \dot{a}}{b a} \label{epsilon_0_equation_3} \, .
\end{align}
In terms of $H$ and $\sigma$ the equations take the form
\begin{align*}
    \dot{H} &= -3 - H^2 \, , \\
    \dot{\sigma} &= 1 - H \sigma \, , \\
   0  &= 1 + \frac{5}{16} H^2 - \frac{3}{16} \sigma^2 - \frac{1}{8} H \sigma  \, .
\end{align*}
The constraint implies that the state space is restricted to two hyperbolic curves. Solving it gives
\begin{align*}
    \sigma_{1,2} =  - \frac{1}{3} H \pm 4 \sqrt{H^2+3} \, .
\end{align*}
Hence, the state space for $\epsilon = 0$ is restricted to the two curves defined by the previous equation. We can even solve the differential equation for $H$ explicitly and obtain
\begin{align*}
    H(t) = - \sqrt{3} \tan\left(\sqrt{3}(c_1+t)\right) \, ,
\end{align*}
with $c_1$ being the integration constant. Inserting the solution for $H$ into the equation for $\sigma$ yields
\begin{align*}
    \dot{\sigma} &= 1 + \sigma  \sqrt{3} \tan\left(\sqrt{3}(c_1+t)\right) \, ,
\end{align*}
which has indeed an explicit solution given by
\begin{align*}
    \sigma =  \frac{c_2}{\cos\left(\sqrt{3} (c_1+t) \right)}- \frac{1}{\sqrt{3}} \tan\left(\sqrt{3} (c_1+t) \right) \, .
\end{align*}
The solutions fulfil the constraint equation. Inserting into it yields
\begin{align*}
    \left( 3 c_2^2 - 16 \right)\frac{1}{\cos\left(\sqrt{3} (c_1+t) \right)^2} = 0 \, .
\end{align*}
Hence $c_2 = \pm 4 \frac{1}{\sqrt{3}}$. The constant $c_1$ has to be determined by the initial conditions. It is then easy to see that the solutions have a finite existence time and the mean curvature takes on all real values. Using the expressions for $H$ and $\sigma$ one can transform back to the original variables $a$ and $b$. But, as the mean curvature takes on all real values we can conclude that the spacetime recollapses. The singularities are again crushing singularities. Hence we have shown the second auxiliary lemma:
\begin{lem}\label{epsilon_0_recollapse}
All solutions to the system (\ref{epsilon_0_equation_1}), (\ref{epsilon_0_equation_2}),(\ref{epsilon_0_equation_3}) recollapse and their mean curvature takes on all real values, hence producing crushing singularities. The asymptotic behavior of solutions is divided into two classes:
\begin{enumerate}
\item Solutions whose initial data is in the set $\sigma_N = O_4(H_N)$are future asymptotic to (\ref{assymptotic_others}) and past asymptotic to (\ref{pancake_solution}). The future singularity is of Kasner type in the sense that one scale factor vanishes while the other diverges. Additionally, the future singularity is also a curvature singularity as the Kretschmann scalar diverges.
\item Solutions whose initial data is in the set $\sigma_N = O_5(H_N)$ are future asymptotic to (\ref{pancake_solution}) and past asymptotic to (\ref{assymptotic_others}). The past singularity is of Kasner type in the sense that one scale factor vanishes while the other diverges. Additionally, the past singularity is also a curvature singularity as the Kretschmann scalar diverges.
\end{enumerate}
\end{lem}

\subsubsection{Curvature $\varepsilon=1$}
Finally, we consider the case $\epsilon = 1$ (spherical symmetry) and proceed in a similar manner. From the equations (\ref{evolution_a_2}),(\ref{evolution_b}) and (\ref{constraint_ab}) we get
  \begin{align}
      \Ddot{b} =& - \frac{\dot{b}^2}{b} - |\Lambda| b - \frac{1}{b} - \frac{\dot{a} \dot{b}}{a} \, , \label{epsilon_1_equation_1} \\
       \Ddot{a} =& - |\Lambda| a - 2 \frac{\dot{a} \dot{b}}{b} \, , \label{epsilon_1_equation_2}   \\
       0 =& \,|\Lambda| + \frac{1}{b^2} + \frac{ \dot{b}^2}{b^2} + 2 \frac{\dot{b} \dot{a}}{b a}  \, . \label{epsilon_1_equation_3}
  \end{align}
Performing the variable transformation to $H$ and $\sigma$ we obtain 
\begin{align*}
    \dot{H} &= -1 - \frac{1}{4} H \sigma - \frac{3}{8} \left( H^2 + \sigma^2\right) \, ,  \\
    \dot{\sigma} &= -1 - \frac{3}{4} H \sigma - \frac{5}{8} H^2 + \frac{3}{8} \sigma^2 \, , \\
    &1 + \frac{5}{16} H^2 - \frac{3}{16} \sigma^2 - \frac{1}{8} H \sigma = - e^{-2y} < 0 \, .
\end{align*}
This system is similar to the system (\ref{evolution_H}), (\ref{evolution_sigma}) and (\ref{constrain_inequality}). The only difference is the constraint inequality, which forces the left-hand side to be smaller than zero rather than larger. Hence the same analysis carries over as already done for the case of hyperbolic symmetry ($\epsilon=-1$). Furthermore, as shown previously, $H$ has to be monotonically decreasing and can be bounded by
\[ H < - C_1 \tan(t - C_2 ) \, ,\] which shows that the existence interval is finite and the function $H$ will take on all real values.

\begin{figure}
    \centering
    \begin{subfigure}[b]{0.3\textwidth}
        \includegraphics[width=\textwidth]{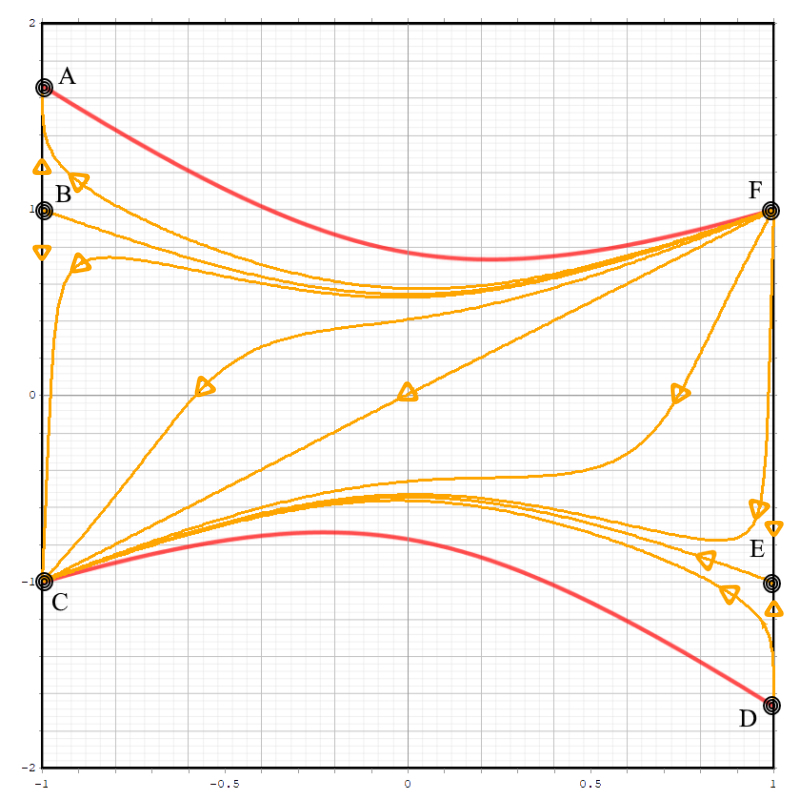}
        \caption{Compactified state space for $\epsilon=-1$.}
        \label{figure_phasespace_bianch_iii_compact}
    \end{subfigure}
    \begin{subfigure}[b]{0.3\textwidth}
        \includegraphics[width=\textwidth]{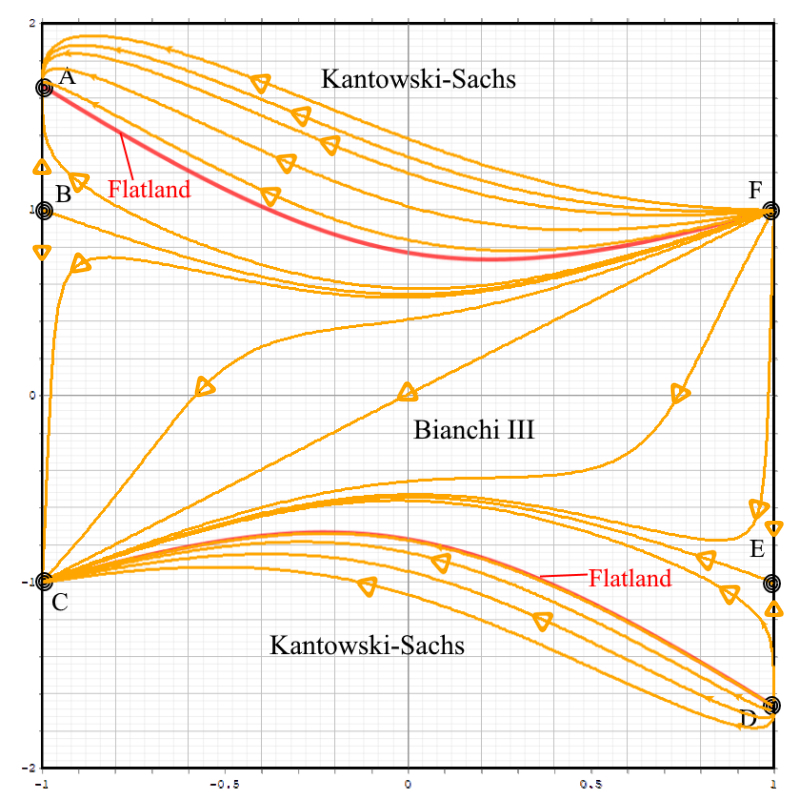}
        \caption{Compactified state space for $\epsilon=1,0,-1$.}
        \label{figure_compactified_phase_space_full}
    \end{subfigure}
    \begin{subfigure}[b]{0.3\textwidth}
          \includegraphics[width=\textwidth]{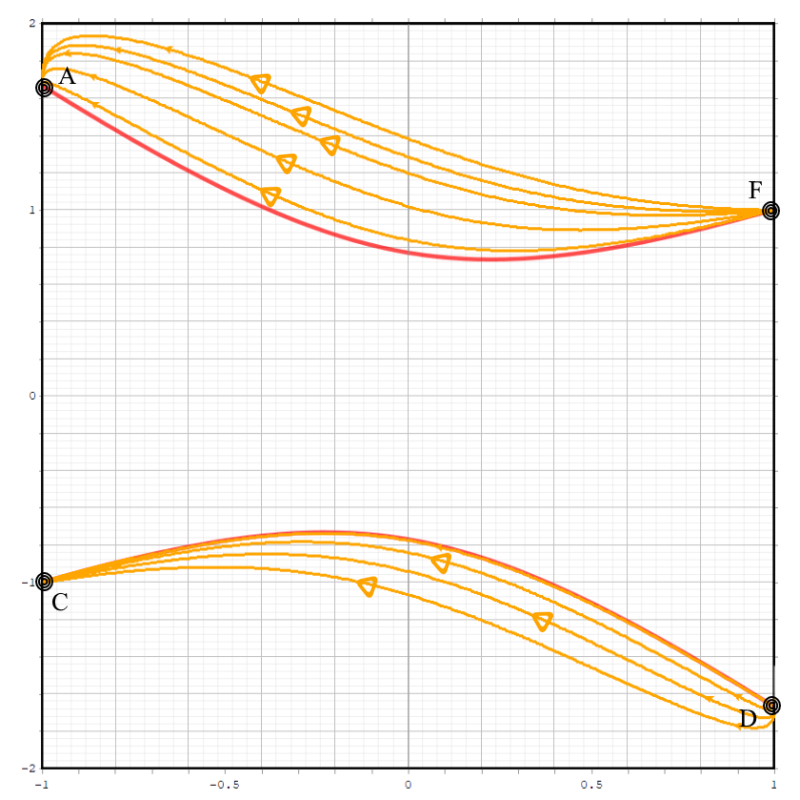}
          \caption{Compactified state space for  $\epsilon = 1$.}
         \label{figure_phasespace_kantowski_sachs_compact}
    \end{subfigure}
    \caption{State space diagrams for $\epsilon=1,0,-1$. $H$ is drawn on the horizontal axis and $\sigma$ is drawn on the vertical axis. Time flows from right to left.}
\end{figure}

State space can also be compactified. This works analogously as in the case of hyperbolic symmetry ($\epsilon = -1$), because the equations are identical. The only difference is the constraint inequality. Instead of being enclosed between the curves $O_4$ and $O_5$ the components of state space are located outside this region. In particular, state space is divided into two invariant sets,
\begin{align*}
    AF:= \left\{ (H_N, \sigma_N) \in S \; | \; O_4(H_N) < \sigma_N \right \} \, , \\
    CD:= \left\{ (H_N, \sigma_N) \in S \; | \; O_5(H_N) > \sigma_N \right \} \, .
\end{align*}
In Figure \ref{figure_phasespace_kantowski_sachs_compact} the compactified state space is drawn qualitatively. In this case the solutions have to lie outside the region which is enclosed by the red curves. There are no saddle points, hence there cannot be any heteroclinic cycles. The sinks $A$, $C$ and the sources $D$, $F$ determine the asymptotic behavior of the flow. The singularities are again crushing singularities as the mean curvature takes on all real values. Hence, we have shown the third and last auxiliary lemma:
\begin{lem}\label{epsilon_1_recollapse}
All solutions to the system (\ref{epsilon_1_equation_1}), (\ref{epsilon_1_equation_2}),(\ref{epsilon_1_equation_3}) recollapse and their mean curvature takes on all real values, hence producing crushing singularities. The asymptotic behavior of solutions is divided into two classes:
\begin{enumerate}
\item Solutions whose initial data lies in the set $AF$ are future asymptotic to (\ref{assymptotic_others}) and past asymptotic to (\ref{pancake_solution}). The future singularity is of Kasner type in the sense that one scale factor vanishes while the other diverges. Additionally, the future singularity is also a curvature singularity as the Kretschmann scalar diverges.
\item Solutions whose initial data lies in the region $CD$ are future asymptotic to (\ref{pancake_solution}) and past asymptotic to (\ref{assymptotic_others}). The past singularity is of Kasner type in the sense that one scale factor vanishes while the other diverges. Additionally, the past singularity is also a curvature singularity as the Kretschmann scalar diverges.
\end{enumerate}
\end{lem}

To sum up, we give a picture of the full state space shown in Figure \ref{figure_compactified_phase_space_full}. Due to the constraint equation state space can be divided into four parts: The lower part is populated by Kantowski-Sachs spacetimes. The middle part is populated by Bianchi type III and the upper part is again Kantowski-Sachs. The two curves that divide Kantowski-Sachs and Bianchi III are the "flatland" of Bianchi type I ($\epsilon = 0$).

\section{The Einstein-scalar field system with $\Lambda<0$}\label{section_curvature_singularity}
As we have seen, the singularities of the pancake solution given by (\ref{pancake_solution}) are crushing singularities but not necessarily curvature singularities. It might a priori be possible to extend spacetime beyond those singularities. We construct related solutions, which do exclusively have curvature singularities by adding a necessarily nontrivial spatially homogeneous massless scalar field. In particular, we will show the following theorem:
\begin{thm}
All spatially homogeneous anisotropic spacetimes of the form
$(F\times \mathbb S^1,a_0^2dz^2+b_0^2\sigma,c_0\sigma+d_0dz^2)$, with $\Lambda < 0$ and a non-vanishing massless scalar field possessing a CMC hypersurface of topology $\mathbb S^1 \times F$, with either $F = \mathbb S^2$, $F=H^2$ or $F=\mathbb S^1 \times \mathbb S^1$ admit a CMC foliation. The mean curvature of the leaves takes on all real values. The spacetimes have future and past curvature singularities as the Kretschmann scalar diverges.
\end{thm}
\begin{proof}
The case $F = \mathbb S^2$ is shown by Lemma \ref{epsilon_1_scalar_recollapse}, $F = \mathbb S^1 \times \mathbb S^1$ is shown by Lemma \ref{epsilon_0_scalar_recollapse} and the $F = H^2$ is shown by Lemma \ref{epsilon_-1_scalar_recollapse}.
\end{proof}
What is left is to show the three auxiliary lemma used in the proof. We start with setting up the equations. The scalar field $\phi$ only depends on the time variable, hence the field equation becomes
\begin{align}
   \nabla^{\mu} \nabla_{\mu} \phi =  g^{\mu \nu} \partial_{\mu} \partial_{\nu} \phi - g^{\mu \nu} \Gamma^{\alpha}_{\mu \nu} \partial_{\alpha} \phi = - \Ddot{\phi} - \frac{\dot{a}}{a} \dot{\phi} - 2 \frac{\dot{b}}{b} \dot{\phi} = 0 \, . \label{scalar_field_equation_2}
\end{align}
We recall the definition of the energy momentum tensor
\begin{align*}
    T_{\mu \nu} = \nabla_{\mu} \phi  \nabla_{\nu} \phi - \frac{1}{2} g_{\mu \nu}   g^{\alpha \beta} \nabla_{\alpha} \phi  \nabla_{\beta} \phi \, .
\end{align*}
The energy momentum tensor yields the following matter terms in the Einstein equations.
\begin{align*}
    \rho &= T_{00} = \dot{\phi}^2 - \frac{1}{2} g_{00}  g^{00} \dot{\phi}^2 = \dot{\phi}^2 - \frac{1}{2}  \dot{\phi}^2 = \frac{1}{2} \dot{\phi}^2 \, , \\
    j &= - T_{01} = \frac{1}{2} g_{01} g^{00}\dot{\phi}^2 = 0 \, , \\
    S_{11} &= T_{11} =  -\frac{1}{2} g_{11}  g^{00} \dot{\phi}^2 = \frac{1}{2} a^2 \dot{\phi}^2 \, ,\\
     S_{22} &=  T_{22} = -\frac{1}{2} g_{22}  g^{00} \dot{\phi}^2 = \frac{1}{2} b^2 \dot{\phi}^2 \, , \\
     S_{33} &=   T_{33} =  -\frac{1}{2} g_{33}  g^{00} \dot{\phi}^2 = \frac{1}{2} b^2 f(\theta)_\epsilon^2 \dot{\phi}^2 \, , \\
     \si{tr}(S) &= S_{11} g^{11} +  S_{22} g^{22} +  S_{33} g^{33} = \frac{3}{2} \dot{\phi}^2 \, .
\end{align*}
Then we calculate the evolution equations and discover that only the constraint equation (\ref{constraint_ab}) is modified by a term depending on $\dot{\phi}$. They read
  \begin{align}
      \Ddot{b} =& - \frac{\dot{b}^2}{b} - |\Lambda| b - \frac{\epsilon}{b} - \frac{\dot{a} \dot{b}}{a} \, , \label{evolution_b_scalar} \\
       \Ddot{a} =& - |\Lambda| a - 2 \frac{\dot{a} \dot{b}}{b} \, ,  \label{evolution_a_2_scalar} \\
       4 \pi \dot{\phi}^2 =& |\Lambda| + \frac{\epsilon}{b^2} + \frac{ \dot{b}^2}{b^2} + 2 \frac{\dot{b} \dot{a}}{b a} \label{constraint_ab_scalar} \, .
  \end{align}
\subsubsection{Curvature $\varepsilon=-1$}
For simplicity, we set $|\Lambda| = 1$ and first study the case $\epsilon = -1$. We introduce  $H:= \frac{\dot{a}}{a}+ 2  \frac{\dot{b}}{b}$, $\sigma:=  \frac{\dot{a}}{a}- 2  \frac{\dot{b}}{b}$, $\Phi:= \dot{\phi}$. The evolution equations become
\begin{align}
    \dot{H} &= -1 - \frac{1}{4} H \sigma - \frac{3}{8} \left( H^2 + \sigma^2\right) - 8 \pi \Phi^2 \, , \label{evolution_H_phi} \\
    \dot{\sigma} &= -1 - \frac{3}{4} H \sigma - \frac{5}{8} H^2 + \frac{3}{8} \sigma^2 + 8 \pi \Phi^2\, . \label{evolution_sigma_phi}
\end{align}
The constraint becomes
\begin{align}
    1 + \frac{5}{16} H^2 - \frac{3}{16} \sigma^2 - \frac{1}{8} H \sigma - 4 \pi \Phi^2 =  e^{-2y} > 0 \, , \label{constrain_inequality_phi}
\end{align}
and the equation for the scalar field becomes
\begin{align}
    \dot{\Phi} = -H \, \Phi\, . \label{equation_phi_with_H}
\end{align}
We compactify the system by performing a reparametrisation according to
\begin{align*}
    H_N:= \frac{H}{N} \, , \quad   \sigma_N:= \frac{\sigma}{N}  \,. \quad \Phi_N:= \frac{\Phi}{N} 
\end{align*}
This casts the evolution equations into the form
\begin{align}
    \dot{H}_N =& \left( -24 \pi \Phi_N^{2}-\frac{9}{8} \left( \sigma_N^{2}+H_N^{2}\right)-\frac{3}{4} \sigma_N H_N-\frac{1}{3} (1-H_N^2)\right)  \left( 1-{H_N^{2}}\right) \, ,
\\
   \dot{\sigma}_N =&  \left( 24 \pi \Phi_N^{2}+\frac{9}{8}\sigma_N^{2}-\frac{9}{4} H_N \sigma_N-\frac{1}{3}(1-H_N^{2})-\frac{15}{8}H_N^{2}\right) \\  &+ H_N \sigma_N 
\left( 24 \pi \Phi_N^{2}+\frac{9}{8} \left( \sigma_N^{2}+H_N^{2}\right) +\frac{3}{4}H_N \sigma_N+\frac{1}{3}(1-H_N^{2})\right)  \, , \\
    \dot{\Phi}_N =& \left( 24 \pi \Phi_N^{2}+\frac{9}{8} \left({\sigma_N}^{2}+{H_N}^{2}\right) +\frac{3}{4} H_N \sigma_N+\frac{1}{3}(1-{{H_N}^{2}})\right)  H_N \Phi_N -3H_N \Phi_N \, .
\end{align}
The constraint equation becomes
\begin{align}
    1 + \frac{5}{16} H_N^2 - \frac{3}{16} \sigma_N^2 - \frac{1}{8} H_N \sigma_N - 4 \pi \Phi_N^2 > 0 \,. \label{constrain_phi}
\end{align}
State space is now given by
\[
 S_\phi:= \left\{ (H_N,\sigma_N,\Phi_N) \in \mathbb{R}^3 \mid -1 \le H_N \le 1\right\} \, .
 \]
The existence of periodic orbits can be excluded as $\dot{H}_N<0$ and thus the monotonicity of $H_N$ is fixed to be monotonically decreasing. But a periodic orbit requires that $H_N$ returns to its original value, which is a contradiction to $H_N$ being monotonically decreasing.

Secondly, the system has invariant sets defined by $\Phi_N = 0$ and $\Phi_N > 0$, $\Phi_N < 0$. We define $Z_\phi = \phi$. Differentiating yields 
\begin{align*}
    \dot{Z}_\phi = \dot{\Phi}_N =& \left( 24 \pi \Phi_N^{2}+\frac{9}{8} \left({\sigma_N}^{2}+{H_N}^{2}\right) +\frac{3}{4} H_N \sigma_N+\frac{1}{3}(1-{{H_N}^{2}})\right)  H_N \Phi_N -3H_N \Phi_N \\
    =& \alpha_\phi  \, \Phi_N \, ,
 \end{align*}
with
\begin{align*}
    \alpha_\phi =& \left( 24 \pi \Phi_N^{2}+\frac{9}{8} \left({\sigma_N}^{2}+{H_N}^{2}\right) +\frac{3}{4} H_N \sigma_N+\frac{1}{3}(1-{{H_N}^{2}})\right)  H_N  -3H_N \, .
\end{align*}
 Thus, the $3$-dimensional state space is separated into three regions. Solving the constraint inequality (\ref{constrain_phi}) defines a hyperboloid in state space, given by \[
 \Omega_0:= \left\{ (H_N,\sigma_N,\Phi_N) \in S_{\phi} \mid 1 + \frac{5}{16} H_N^2 - \frac{3}{16} \sigma_N^2 - \frac{1}{8} H_N \sigma_N - 4 \pi \Phi_N^2 = 0\right\} \, .
 \]
All orbits for the case $\epsilon = -1$ lie inside the hyperboloid $\Omega_0$. We define the following invariant sets:
\begin{align*}
    \Omega^{+}_{-1}:=& \left\{ (H_N,\sigma_N,\Phi_N) \in S_{\phi} \mid  \phi_N>0 \, , \;  1 + \frac{5}{16} H_N^2 - \frac{3}{16} \sigma_N^2 - \frac{1}{8} H_N \sigma_N - 4 \pi \Phi_N^2 < 0 \right\} \, ,\\
    \Omega^{-}_{-1}:=& \left\{ (H_N,\sigma_N,\Phi_N) \in S_{\phi} \mid  \phi_N<0 \, , \;  1 + \frac{5}{16} H_N^2 - \frac{3}{16} \sigma_N^2 - \frac{1}{8} H_N \sigma_N - 4 \pi \Phi_N^2 < 0 \right\} \, ,\\
    \Omega^{+}_1:=& \left\{ (H_N,\sigma_N,\Phi_N) \in S_{\phi} \mid  \phi_N>0 \, , \;   1 + \frac{5}{16} H_N^2 - \frac{3}{16} \sigma_N^2 - \frac{1}{8} H_N \sigma_N - 4 \pi \Phi_N^2 > 0 \right\} \, , \\
        \Omega^{-}_1:=& \left\{ (H_N,\sigma_N,\Phi_N) \in S_{\phi} \mid  \phi_N<0 \, , \;   1 + \frac{5}{16} H_N^2 - \frac{3}{16} \sigma_N^2 - \frac{1}{8} H_N \sigma_N - 4 \pi \Phi_N^2 > 0 \right\} \, .
\end{align*}

Searching for fixed points provides the same fixpoints as in the case with a vanishing scalar field, we label them similarly. Hence, the results from Section \ref{section_vacuum}  describe the plane $\Phi_N = 0$ completely. Additionally we obtain two normally hyperbolic sets of fixed points, parameterised by the curves\begin{align*}
    &\gamma_{E_1}(s) = \left(1,s, f_1(s) \right)\, ,&  \quad &\gamma_{E_2}(s) = \left(1,s,-f_1(s) \right)\, ,&  \\
  &\gamma_{B_1}(s) = \left(-1,s,f_2(s)\right)\, ,&  \quad &\gamma_{B_2}(s)= \left(-1,s,-f_2(s)\right)\, ,& 
\end{align*}
with auxiliary functions
\begin{align*}
 f_1(s)&:=\frac{1}{8\sqrt{\pi}} \sqrt{-3 s^2 - 2 s + 5}\, , \;
&\si{for} \; & - \frac{5}{3} \le s \le 1 \, ,&  \\
 f_2(s)&:=\frac{1}{8\sqrt{\pi} }\sqrt{-3 s^2 + 2 s + 5}\, , \;
&\si{for} \; & -1 \le s \le \frac{5}{3}\, . & 
\end{align*}
And four single fixed points
\begin{align*}
    E_1 = (1,-1,\frac{1}{4 \sqrt{\pi}}) \, , \hspace{0.25cm} E_2 = (1,-1,-\frac{1}{4 \sqrt{\pi}}) \, , 	\hspace{0.5cm}
        B_1 = (-1,1,\frac{1}{4 \sqrt{\pi}}) \, , \hspace{0.25cm} B_2 = (-1,1,-\frac{1}{4 \sqrt{\pi}}) \, .
\end{align*}
A picture of the state space is drawn in Figure \ref{figure_compactified_phase_space_full}. Linearization and calculation of the eigenvalues and eigenvectors of the fixed points which lie on the plane $\Phi_N = 0$ gives that they all have an additional eigenvector $(0,0,1)$ pointing in direction of $\Phi_N$ and an additional eigenvalue which vanishes. Only the fixed points $E$ and $B$ have a non vanishing eigenvalue in $\Phi_N$ direction, namely $\lambda_{3B} = -3$ and $\lambda_{3E} = 3$. The other eigenvalues are shown in Table \ref{tab_fixed_points}.

\begin{table}[ht]
    \centering
    \begin{tabular}{lr}
    
    \begin{tabular}{c|c|c|c}
    Fixed point & $\lambda_1$ & $\lambda_2$ & $\lambda_3$ \\
    \hline
    $E_1$ & 0 & 3 & 6 \\
    \hline
    $E_2$ & 0 & 3 & 6 \\
    \hline
    $B_1$ & -6 & -3 & 0 \\
    \hline
    $B_2$ & -6 & -3 & 0 \\
    \end{tabular}
    &\quad
    \begin{tabular}{c|c|c|c}
    Fixed point & $\lambda_1$ & $\lambda_2$ & $\lambda_3$ \\
    \hline
    $\gamma_{E_1}$ & $\nicefrac{3}{2}\, s - \nicefrac{9}{2}$ &-6 & 0 \\
    \hline
    $\gamma_{E_2}$ & $\nicefrac{3}{2}\, s - \nicefrac{9}{2}$ &-6 & 0 \\
    \hline
    $\gamma_{B_1}$ & $\nicefrac{3}{2}\, s + \nicefrac{9}{2}$ &0 & 6 \\
    \hline
    $\gamma_{B_2}$ & $\nicefrac{3}{2}\, s + \nicefrac{9}{2}$ &0 & 6  \\
    \end{tabular}
    \end{tabular}
    \vspace{0.5cm}
    \caption{Eigenvalues at the fixed points with a non-vanishing scalar field.}
    \label{tab_fixed_points}
\end{table}
Calculating the eigenvectors shows that all fixed points in the $\Phi_N=0$ plane get an additional eigenvector $(0,0,1)$ pointing in $\Phi_N$ direction. 

All fixed points except the hyperbolic ones ($B,E$) have a one dimensional center manifold. The saddle point $B$ has a two dimensional unstable manifold which is given by the boundary of state space. Additionally it has a one dimensional stable manifold given by the orbit $O_2$ as defined in the previous section. The saddle point $E$ has a two dimensional stable manifold which is given by the boundary of state space. Furthermore it has a one dimensional unstable manifold given by the orbit $O_3$. 

Now we want to find the one dimensional center manifolds of the other fixed points. For that we need the eigenvectors of the fixed points 
$E_1,E_2,B_1,B_2$ and the fixed point sets for the vanishing eigenvalue as they form a basis of the tangential space of the center manifold at the respective fixed point. They are given by
\begin{align*}
        \Vec{\lambda_3}_{E_1} &= \begin{pmatrix}
 1 \\
 0  \\
 - \frac{5}{72 \sqrt{\pi}}
\end{pmatrix} \, , &\;
\Vec{\lambda_3}_{E_2} &= \begin{pmatrix}
 1 \\
 0  \\
  \frac{5}{72 \sqrt{\pi}}
\end{pmatrix}  \, , &\;
        \Vec{\lambda_3}_{B_1} &= \begin{pmatrix}
 0 \\
 1  \\
 \frac{1}{8 \sqrt{\pi}}
\end{pmatrix} \, , &\;
\Vec{\lambda_3}_{B_2} &= \begin{pmatrix}
 0 \\
 1  \\
 - \frac{1}{8 \sqrt{\pi}}
\end{pmatrix}   \, ,
\\
\Vec{\lambda_3}_{\gamma_{E_1}} &= \begin{pmatrix}
 0 \\
 1  \\
 g_1(s)
\end{pmatrix} \, ,  &\;
\Vec{\lambda_3}_{\gamma_{E_2}} &= \begin{pmatrix}
 0 \\
 1  \\
 -g_1(s)
\end{pmatrix}  \, ,  &\;
        \Vec{\lambda_3}_{\gamma_{B_1}} &= \begin{pmatrix}
 0 \\
 1  \\
 g_2(s)
\end{pmatrix}  \, ,  &\;
\Vec{\lambda_3}_{\gamma_{B_2}} &= \begin{pmatrix}
 0 \\
 1  \\
 -g_2(s)
\end{pmatrix}  \, ,
\end{align*}
where we define the auxiliary functions \begin{align*}
g_1(s)&:=-\frac{\left( 3 s+1\right)}{ 8\sqrt{\pi }  \sqrt{-3 {{s}^{2}}-2 s+5}}\, , \;
&\si{for} \; &- \frac{5}{3} \le s \le 1 \; \si{and} \; s \neq -1 \, ,&  \\
g_2(s)&:= -\frac{\left( 3 s-1\right)}{8\sqrt{\pi }\sqrt{-3 {{s}^{2}}+2 s+5}}\, , \;
&\si{for} \; &  -1 \le s \le \frac{5}{3}  \; \si{and}\; s \neq 1 \, .& 
\end{align*}
We note that
\[ \gamma_{E_1}(-1) = E_1 \, , \quad \gamma_{E_2}(-1) = E_2 \, , \quad \gamma_{B_1}(1) = B_1 \, , \quad \gamma_{B_2}(1) = B_2 \, ,\]
which means that $E_1$, $E_2$, $B_1$ and $B_2$ are already part of the fixed point sets. But for the parameter values $s=1$ and $s=-1$ the eigenvectors of $\gamma_{E_1}$ ,  $\gamma_{E_2}$ and $\gamma_{B_1}$ ,  $\gamma_{B_2}$ are not defined. However, the eigenvectors of $E_1$, $E_2$ and  $B_1$, $B_2$ are well defined.

It is easy to see that  \[f_1'(s) = g_1(s) \, , \quad f_2'(s) = g_2(s)\] and thus \[
\gamma_{E_1}'(s) =  \Vec{\lambda_3}_{\gamma_{E_1}} \, , \; \gamma_{E_2}'(s) =  \Vec{\lambda_3}_{\gamma_{E_2}} \, , \; \gamma_{B_1}'(s) =  \Vec{\lambda_3}_{\gamma_{B_1}} \, , \; \gamma_{B_2}'(s) =  \Vec{\lambda_3}_{\gamma_{B_2}} \, .
\]
Thus, the center manifolds of the fixed points $E_1,E_2,\gamma_{E_1},\gamma_{E_2},F,D$ are precisely given by the 1-dimensional manifold defined by the union of  the curves $\gamma_{E_1}$ and $\gamma_{E_2}$.  Similarly, the center manifolds of the fixed points $B_1,B_2,\gamma_{B_1},\gamma_{B_2},A,C$ are precisely given by the 1-dimensional manifold defined by the union of the curves $\gamma_{B_1}$ and $\gamma_{B_2}$. 

Furthermore, $E_1,E_2,\gamma_{E_1},\gamma_{E_2}$ have 2-dimensional stable manifolds, while $B_1,B_2,\gamma_{B_1},\gamma_{B_2}$ have 2-dimensional unstable manifolds.

\begin{figure}[th]
    \centering
    \includegraphics[width=0.5\textwidth]{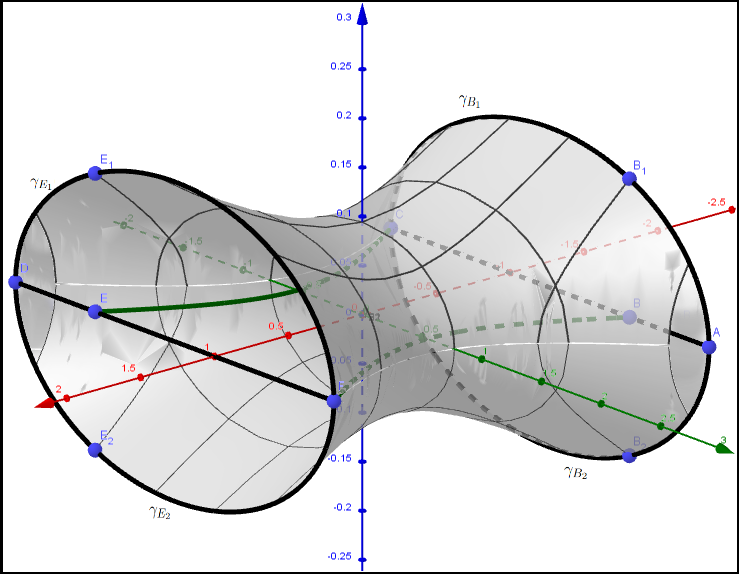}
    \caption{Picture of compactified state space. $H_N$ is drawn on the red axis and $\sigma_N$ is drawn on the green axis and $\phi_N$ is drawn on the red axis.}
    \label{figure_compactified_phase_space_full}
\end{figure}

This explains the behavior of solutions completely. Due to Theorem \ref{fix_point_sets} no solution can flow into the fixed points in the $\Phi_N=0$ plane, Hence, solutions with initial data $\Phi_N>0$ will also have $\Phi_N>0$ in their asymptotic past or future. Solutions with initial data in the invariant set $\Omega^{+}_{-1}$ are future asymptotic to $\gamma_{E_1}$ and past asymptotic to $\gamma_{B_1}$, while solutions with initial data in the invariant set $\Omega^{-}_{-1}$ are future asymptotic to $\gamma_{E_2}$ and past asymptotic to $\gamma_{B_2}$.

What is left is to calculate the asymptotic solutions of the fixed point sets. We will do this for the set $\gamma_{E_1}$, the other cases work analogously.
We have that
\begin{align*}
    s \, H_N = \sigma_N \, , \quad   \frac{1}{f_1(s)} \Phi_N =  \, H_N \, ,
\end{align*}
and thus
\begin{align*}
    s \, H = \sigma \, , \quad \frac{1}{f_1(s)} \Phi =  \, H \, .
\end{align*}
Hence we have from the evolution equation for $H$ that
\begin{align*}
    \dot{H} = - 1 - \left(\frac{3}{8}+ \frac{1}{4} s + \frac{3}{8} s^2 + 8 \pi f_1(s)^2\right) \, H^2.
\end{align*}
The term \[G_1(s):=\frac{3}{8}+ \frac{1}{4} s + \frac{3}{8} s^2 + 8 \pi f_1(s)^2\] is strictly positive.
Hence we have
\begin{align}
    H(t) =- \frac{1}{\sqrt{G_1(s)}} \tan\left( \sqrt{G_1(s)} (t+C_1) \right) \, ,
\end{align}
with $C_1$ being some integration constant. We also have the relations
\[\frac{\dot{a}}{a} = 2 \frac{1+s}{1-s}\frac{\dot{b}}{b} \, , \quad H = 2 \left(1 + \frac{1+s}{1-s} \right) \frac{\dot{b}}{b} \, ,\] which only gives a meaningful result for $s \neq 1$. The case $s=1$ corresponds to the fixed point $F$ from the previous senction. But this is not relevant for this analysis. In this case we get for the asymptotic solution of $\gamma_{E_1}$.  Hence we get
\begin{align}
    b_{\gamma_{E_1}}(t) = C_2 \cos^{r_1}\left( \sqrt{G_1(s)} (t+C_1) \right) \, ,  \label{assymptotic_gammae1}
\end{align}
with $r_1 = 2  \left(1 + \frac{1+s}{1-s} \right) G_1(s) \, .$
Due to symmetry  $\gamma_{E_2}$ has the same solution given by (\ref{assymptotic_gammae1}). The single fixed points  $E_1$ and $E_2$ have the same solution for parameter value $s=-1$.

For the set $\gamma_{B_1}$ we have the relations
\[\frac{\dot{a}}{a} = 2 \frac{1-s}{1+s}\frac{\dot{b}}{b} \, , \quad H = 2 \left(1 + \frac{1-s}{1+s} \right) \frac{\dot{b}}{b} \, ,\] which only gives a meaningful result for $s \neq -1$. The case $s=-1$ corresponds to the fixed point $C$ from the previous senction. But this is not relevant for this analysis. And solving gives 
\begin{align}
    b_{\gamma_{B_1}}(t) = C_2 \cos^{r_2}\left( \sqrt{G_2(s)} (t+C_1) \right) \, ,  \label{assymptotic_gammae2}
\end{align}
with $r_2 =  2  \left(1 + \frac{1-s}{1+s} \right) G_2(s)$ and 
\[G_2(s):=\frac{3}{8} - \frac{1}{4} s + \frac{3}{8} s^2 + 8 \pi f_2(s)^2 \, .\]
Again, due to symmetry  $\gamma_{B_2}$ has the same solution given by (\ref{assymptotic_gammae1}). The single fixed points  $B_1$ and $B_2$ have the same solution for parameter value $s=1$.

In particular, for all the asymptotic solutions $b$ vanishes at two points in time, one in the future and one in the past. When $b$ vanishes the Kretschmann scalar diverges and a curvature singularity occurs.

We have shown the following lemma:
\begin{lem}\label{epsilon_-1_scalar_recollapse}
    All solutions to the system (\ref{scalar_field_equation_2}), (\ref{evolution_b_scalar}), (\ref{evolution_a_2_scalar}), (\ref{constraint_ab_scalar}) for $\epsilon=-1$ recollapse and their mean curvature takes on all real values. Additionally, the Kretschmann scalar diverges when the singularities are approached, producing future and past curvature singularities.
\end{lem}
\subsubsection{Curvature $\varepsilon=1$}
The case $\epsilon = 1$ works analogously. The evolution equations are identical, only the constraint equation is different. In this case it reads
\begin{align*}
    1 + \frac{5}{16} H_N^2 - \frac{3}{16} \sigma_N^2 - \frac{1}{8} H_N \sigma_N - 4 \pi \Phi_N^2 > 0 \, .
\end{align*}
Hence, solutions with initial data in the invariant set $\Omega^{+}_{1}$ are future asymptotic to $\gamma_{E_1}$ and past asymptotic to $\gamma_{B_1}$. While solutions with initial data in the invariant set $\Omega^{-}_{1}$ are future asymptotic to $\gamma_{E_2}$ and past asymptotic to $\gamma_{B_2}$. We obtain the following theorem:
\begin{lem}\label{epsilon_1_scalar_recollapse}
    All solutions to the system (\ref{scalar_field_equation_2}), (\ref{evolution_b_scalar}), (\ref{evolution_a_2_scalar}), (\ref{constraint_ab_scalar}) for $\epsilon=1$ recollapse and their mean curvature takes on all real values. Additionally, the Kretschmann scalar diverges when the singularities are approached, producing future and past curvature singularities.
\end{lem}
\subsubsection{Curvature $\varepsilon=0$}
In the case $\varepsilon = 0$ we have 
\begin{align*}
    1 + \frac{5}{16} H_N^2 - \frac{3}{16} \sigma_N^2 - \frac{1}{8} H_N \sigma_N - 4 \pi \Phi_N^2 = 0 
\end{align*} as the constraint equation. Hence those solutions are all in the set $\Omega_0$. As $\Phi_N=0$, $\Phi_N>0$ and $\Phi_N<0$ are all invariant sets, dividing the set of solutions into three classes. The case $\Phi_N=0$ has already been covered previously. For the other cases we define the sets:
\begin{align*}
    \Omega^{+}_{0}:= \left\{ (H_N,\sigma_N,\Phi_N) \in \Omega_0 \mid  \phi_N>0    \right\} \, ,\quad
    \Omega^{-}_{0}:= \left\{ (H_N,\sigma_N,\Phi_N) \in \Omega_0 \mid  \phi_N<0    \right\} \, .
\end{align*}
Hence, solutions with initial data in the invariant set $\Omega^{+}_{0}$ are future asymptotic to $\gamma_{E_1}$ and past asymptotic to $\gamma_{B_1}$. Solutions with initial data in the invariant set $\Omega^{-}_{0}$ are future asymptotic to $\gamma_{E_2}$ and past asymptotic to $\gamma_{B_2}$. Thus, we obtain the following lemma:
\begin{lem}\label{epsilon_0_scalar_recollapse}
    All solutions to the system (\ref{scalar_field_equation_2}), (\ref{evolution_b_scalar}), (\ref{evolution_a_2_scalar}), (\ref{constraint_ab_scalar}) for $\epsilon=0$ recollapse and their mean curvature takes on all real values. Additionally, the Kretschmann scalar diverges when the singularities are approached, producing future and past curvature singularities.
\end{lem}

\section{Remarks on open sets of initial data}\label{section_remarks}
Hawking's singularity theorem allows to conclude incompleteness of a spacetime under relatively general conditions. It has been pointed out in \cite{ringstrom} that it can be used to show recollapse for an open neighborhood of initial data, in particular for initial data without symmetries. The Theorem can be found for instance in \cite[page 431]{neill} it can be paraphrased as follows.

\begin{lem}
A globally hyperbolic spacetime with $\mathrm{Ric}(V,V)\geq 0$ for every timelike vector $V$, which has two Cauchy hypersurfaces of strictly positive and negative mean curvature, respectively, is future and past timelike geodesically incomplete.
\end{lem}

\begin{rem}
Hawking's singularity theorem applies to Einsteinian spacetimes with a negative cosmological constant. The necessary condition for the theorem to apply is $\mathrm{Ric}(V,V)\geq 0$ for any time-like vector field $V$. The Einstein equations with negative cosmological constant $\Lambda$ imply
\eq{
\mathrm{Ric}(V,V)=-|\Lambda|g(V,V)=|\Lambda||g(V,V)|>0,
}
which implies that the necessary non-negativity of the Ricci-tensor is given and Hawking's theorem applies.
\end{rem}

The following theorem is then an immediate consequence of the local stability of the Einstein flow and the foregoing lemma.

\begin{thm}
Let $(F\times \mathbb S^1,a_0^2dz^2+b_0^2\sigma,c_0\sigma+d_0dz^2)$ be homogeonous initial data with a negative cosmological constant $\Lambda$, where $(F,\sigma)$ is a closed surface of constant curvature. Let $(F\times \mathbb S^1, g_0,k_0)$ be any initial data set with $\Lambda$ such that
\eq{
\|g_0-(a_0^2\sigma+b_0^2dz^2)\|_{H^4}+\|k_0-c_0\sigma+d_0dz^2\|_{H^3}\leq \varepsilon
}
for $\varepsilon>0$ sufficiently small. Then the maximal development of $(F\times \mathbb S^1, g_0,k_0)$ is future and past incomplete.
\end{thm}

For the case of a massless scalar field we have 
\begin{align}
    \si{Ric}(V,V) = -|\Lambda| g(V,V) + 8 \pi T(V,V) = |\Lambda| \,|g(V,V)| + 8 \pi T(V,V) >0 \, ,
\end{align}
as $ T(V,V) > 0$ due to the matter model satisfying the weak energy condition. Again, the local stability property in the presence of a scalar field yields the following theorem.

\begin{thm}
Let $(F\times \mathbb S^1,a_0^2dz^2+b_0^2\sigma,c_0\sigma+d_0dz^2, \phi_0)$ be homogeonous initial data for the Einstein-scalar field system with a negative cosmological constant $\Lambda$, where $(F,\sigma)$ is a closed surface of constant curvature and $\phi_0 \in \mathbb{R}$ is the initial data of the scalar field $\phi \in \mathcal{C}^\infty(\mathbb{R})$. Let $(F\times \mathbb S^1, g_0,k_0,\Phi_0)$ be any initial data set with $\Lambda$ such that
\eq{
\|g_0-(a_0^2\sigma+b_0^2dz^2)\|_{H^4}+\|k_0-c_0\sigma+d_0dz^2\|_{H^3}+|\Phi_0-\phi_0|\leq \varepsilon
}
for $\varepsilon>0$ sufficiently small. Then the maximal development of $(F\times \mathbb S^1, g_0,k_0,\Phi_0)$ is future and past incomplete.
\end{thm}

\section{Discussion}
All known examples of spacetimes solving the Einstein equations with a negative cosmological constant recollapse by developing at least crushing singularities in the past and the future. We have found no counter example to this behavior. We have shown that all spatial homogeneous anisotropic spacetimes, of the topology $F\times \mathbb S^1$ with a negative cosmological constant recollapse and generate crushing singularities in the past and the future, \textbf{independent from their spatial topology}. Furthermore, we have seen that for all possible cases of spatial topology the presence of a massless scalar field in addition causes these singularities to be curvature singularities.



\section*{Appendix}
\subsection{State spaces of ordinary differential equations}
In this section we want to give a very brief overview of dynamical system analysis and the concept of a state space. In the first subsection we will recall the most important theorems which are used later on. The second subsection will explain the procedure of compactification in more detail. We orient ourselves on the books \cite{wainwright1997dynamical} by Wainright and \cite{perko1996differential} by Perko.

\subsection{Dynamical system analysis} \label{subsection_dynamical_systems}
Dynamical system analysis deals with analysing the qualitative behavior of solutions of the $n$-dimensional system
\begin{align}
    \dot{x} = f(x) \, , \label{ordinary_differential_equation}
\end{align}
of ordinary differential equations (ODEs), with $x \in \mathbb{R}^n$ and $f: \mathbb{R}^n \rightarrow \mathbb{R}^n$. The main idea is to view the solutions of (\ref{ordinary_differential_equation}) as states, which are elements of a \textit{state space} (also called \textit{state space}) $X$. In principle, the state space can be the whole of $R^n$. If one has additional constraint equations the state space is restricted according to the constraints. In this picture, the evolution of the states $x$ is governed by the system of ODEs given by (\ref{ordinary_differential_equation}). 

In this context it is convenient to define the \textit{flow} of an ODE.
\begin{dfn}[Flow]
Consider the ODE system given by (\ref{ordinary_differential_equation}), with $f \in \mathcal{C}^1(\mathbb{R}^n)$. Let $\Phi_a(t)$ denote the unique maximal solution satisfying $\Phi_a(0)=a$, for $a \in \mathbb{R}^n$, with existence interval $I=\mathbb{R}$. The flow is defined as one-parameter family of maps $\{ \phi_t\}_{t \in \mathbb{R}}$, mapping $\mathbb{R}^n$ into itself, such that
\[ \phi_t(a) = \Phi_a(t)\] for all $a \in \mathbb{R}^n$.
\end{dfn}
If the existence interval of solutions is finite one can define a  \textit{semi-flow} by replacing $\mathbb{R}$ with the existence interval. The flow has several important properties. In particular, a flow determines an ODE-system on $\mathbb{R}^n$ and vice versa. Additionally, the flow allows a precise definition of a dynamical system.
\begin{dfn}[Dynamical system]
A continuous dynamical system means the flow $\{ \phi_t\}_{t \in \mathbb{R}}$ of a system of ODEs in $\mathbb{R}^n$.
\end{dfn}
With the definitions at hand we want to obtain a qualitative picture of the state space. The points in a state space can be divided into two different types, namely \textit{equilibrium points} (also called \textit{fixed points}) and ordinary points.
\begin{dfn}[Equilibrium Point]
An equilibrium point $x_0$ of the ODE given by (\ref{ordinary_differential_equation}) satisfies $f(x_0) = 0$ or equivalently $\phi_t (x_0) = x_0$, with $x_0 \in \mathbb{R}^n$, for all $t \in \mathbb{R}$.
\end{dfn}
All points which are not equilibrium points are called ordinary points.
\begin{dfn}[Orbit]
Given a flow $\phi_t$ we define the function $\psi(t,x):= \phi_t(x)$, with $x \in \mathbb{R}^n$ and $t \in I  \subset \mathbb{R}$. The \textit{orbit of $x$} (also called \textit{solution curves} or \textit{flow lines}) is defined by $\psi(.,x): I \rightarrow \mathbb{R}^n$.
\end{dfn}
Its worth to note that the orbit through an equilibrium point is the point itself. The importance of this definition is that there exists different types of orbits. If there exists a $t_0$, such that $\phi_{t_0} (x_0) = x_0$ and $t_0>0$, then the associated orbit is called \textit{periodic orbit}. Trivially, all fixed points are also periodic orbits. An orbit connecting equilibrium points is called a \textit{heteroclinic orbit}. An orbit connecting an equilibrium point to itself is called \textit{homoclinic orbit}.

A dynamical system can also have sets which are not changed by the flow (or the corresponding ODE).
\begin{dfn}[Invariant set]
Let $A \subset \mathbb{R}^n$. The set $A$ is called \textit{invariant set} of the flow $\phi_t$ if for all $x \in A$ and for all $t \in \mathbb{R}$, $\phi_t(x) \in A$.
\end{dfn}

For example, invariant sets are single orbits (like equilibrium points or periodic orbits). There exists several theorems which can be used to find invariant sets. We only need the following:
\begin{thm} \label{theorem_invariant_sets}
Given the ODE (\ref{ordinary_differential_equation}) on $\mathbb{R}^n$, with flow $\phi_t$. Let $Z: \mathbb{R}^n \rightarrow \mathbb{R}$ be a $\mathcal{C}^1(\mathbb{R}^n)$ function. If $Z(x,.)$ satisfies \[\frac{\si{d}}{\si{d}t} Z(x,.) = \alpha(x,.) \, Z(x,.) \, ,\] with $\alpha: \mathbb{R}^n \rightarrow \mathbb{R}$ being a continuous function and $x$ being a solution to the ODE, then the regions defined by $Z = 0$, $Z>0$ and $Z<0$ are invariant sets of the flow $\phi_t$.
\end{thm}
\begin{proof}
The proof of the theorem can be found in the book of \cite{wainwright1997dynamical} on page 92.
\end{proof}
Equilibrium points can be classified by linearization of the ODE at the equilibrium point. The linear approximation of the ODE near the equilibrium point $x_0$ yields a new, linear ODE, which is given by
\begin{align}
    \dot{x} \approx Df(x_0) \left( x-x_0\right) \, ,
\end{align}
with
\begin{align*}
   Df(x_0) = \left( \frac{\partial f_i }{\partial x_j}  \right)_{x=x_0} \, .
\end{align*}
The eigenvalues of the matrix $Df(x_0)$ determine the behavior of the solutions near the equilibrium point.
\begin{dfn}[Sinks, Sources and Saddle points]
An equilibrium point $x_0$ of the ODE  (\ref{ordinary_differential_equation}) is called \textit{hyperbolic} if all eigenvalues of $Df(x_0)$ have non-vanishing real part. A hyperbolic equilibrium point is called a \textit{local sink} (\textit{local source}) if all eigenvalues of $Df(x_0)$ have positive non-vanishing real part (negative non-vanishing real part). A hyperbolic equilibrium point which is neither a source or a sink is called \textit{saddle point}.
\end{dfn}

The following theorem says that linearization gives indeed the qualitative picture of the local behavior of the flow near an equilibrium point.
\begin{thm}[Hartman-Grobman] \label{theorem_hartman_grobman}
Given the ODE (\ref{ordinary_differential_equation}). If $x_0$ is a hyperbolic equilibrium point, then there exists a small neighbourhood of $x_0$ on which the flow $\phi_t$ can be mapped with an orientation preserving homeomorphism to the flow of the linearization of the ODE at $x_0$.
\end{thm}
\begin{proof}
The proof can be found in \cite{wainwright1997dynamical} on page 96.
\end{proof}
Given an equilibrium point, the Hartman-Grobman theorem says that all flow lines are in a neighbourhood of the equilibrium point topological equivalent to the flow lines of the linearized system. Hence, linearization can be used to study the solutions of an ODE near the equilibrium points.

To understand the asymptotic behaviour of the solutions of the ODE we have to introduce the concept of $\alpha$- and $\omega$-limit sets.
\begin{dfn}[Limit sets]
Let $\phi_t$ be a flow and let $x_0 \in \mathbb{R}^n$. A point $a \in \mathbb{R}^n$ is an $\omega$-limit point of $x_0$ if there exists a sequence $t_n \rightarrow +\infty$ such that
\[ \lim\limits_{n \rightarrow \infty} \phi_{t_n}(x_0) = a \, .\] The set of all $\omega$-limit points is called $\omega$-limit set. Analogously one defines $\alpha$-limit points and $\alpha$-limit sets by using a sequence $t_n \rightarrow -\infty$.
\end{dfn}
It is worth to note that such limit-sets are invariant sets.
\begin{dfn}[Attractors]
Given a flow $\phi_t$, the future (past) attractor $A^+$ ($A^-$) is the smallest closed invariant set such that $\omega(x) \subset A^+$ ($\alpha(x) \subset A^-$) for all $x \in \mathbb{R}^n$, except for a set of measure zero.
\end{dfn}
For example, consider an ODE with two equilibrium points, one is a source and one is a sink. If one can exclude periodic orbits, then the future attractor of all orbits of the system will be the sink, while the past attractor will be the source.

An important theorem to exclude periodic orbits is the so-called \textit{monotonicity principle}.
\begin{thm}[Monotonicity principle]\label{theorem_monotonicity_principle}
Let $A \subset \mathbb{R}^n$ be an invariant set of the flow $\phi_t$. Let $Z \in \mathcal{C}^1(A)$ and $x$ be a solution of the ODE associated to the flow. Let furthermore $I$ denote the maximal existence interval of the solution. If $Z(\phi_t(x))$ is a monotone decreasing or increasing function of $t \in I$, then $A$ cannot have an equilibrium point, nor periodic orbits or homoclinic orbits.
\end{thm}
\begin{proof}
A proof can be found in \cite{wainwright1997dynamical} on page 93.
\end{proof}
If one considers only a $2$-dimensional system of ODEs, then there exists a strong theorem known by the name Poincare-Bendixson. It can be used to obtain more information about the $\alpha$- and $\omega$-limit sets of a system.
\begin{thm}[Generalized Poincare-Bendixson] \label{theorem_poincare}
Consider the ODE \[ \dot{x} = f(x) \, ,\] with solution $x$, maximal existence interval $I$ and $x(t) \in \mathbb{R}^2$. Furthermore, $f$ is assumed to be $\mathcal{C}^2$ in its arguments. If there exists a finite number of equilibrium points, then any compact limit set is either an equilibrium point, a periodic orbit or the union of equilibrium points and heteroclinic or homoclinic orbits.
\end{thm}
\begin{proof}
The proof can be found in \cite{perko1996differential} on page 245 and following.
\end{proof}

Lastly, we remark what happens if one has a set of fixed points. Then each fixed point of this set necessarily has eigenvalues with vanishing real part. If an $r$-dimensional fixed point set has $r$ eigenvalues with vanishing real part it is called \textit{normally hyperbolic}. In such a case we want to highlight the following theorem from \cite{aulbach1984}, page 36:
\begin{thm}\label{fix_point_sets}
     Let $f: \mathbb{R}^n \rightarrow \mathbb{R}^n$ be three times continuously differentiable and assume that \[ \dot{x} = f(x)\] has a one dimensional normally hyperbolic fixed point set $\gamma$. Then there exists an neighborhood $N \subset \mathbb{R}^n$ of  $\gamma$, such that any solution  having a positive (negative) semi-trajectory in $N$ lies on the stable (unstable) manifold of some fixed point on $\gamma$.
\end{thm}
A positive (negative) semi-trajectory is just the restriction of an orbit to the positive (negative) part of the existence interval.

\subsection{Compactification of a state space} \label{subsection_cumpactification}
For a better understanding of the asymptotic behavior of solutions of a system of ODEs it is convenient to compactify the state space. Roughly speaking one \textit{maps infinity onto one}. The main idea is by performing a suitable transformation to turn asymptotic regions into fixed points, which can be studied by using the techniques described in section \ref{subsection_dynamical_systems}. One can either use a stereographic projection, or in the $2$-dimensional case one can consider the so-called Poincare sphere. In any case, the important point is that such transformations will generate a new state space whose flow lines are homeomorphic to the flow lines of the original state space. For more details we refer the interested reader to \cite{perko1996differential}, page 267 and following. 

In context of cosmology, ODE-systems appear if one considers spatial homogeneous spacetimes, as metrics of such spacetimes will only have coefficients depending on the time variable, but not on the space variables. In this situation the Einstein equations will become a system of second order ODEs. To use dynamical system analysis, one usually defines the Hubble scalar $H$ (which is equal to the mean curvature of the hypersurfaces of the spacetime in question), the shear scalar $\sigma$ and some other quantities usually denoted by $\Omega$, which describe the matter model in question, as new dynamical variables. See for example section \ref{section_curvature_singularity}. Then compactification is usually done by introducing \textit{expansion normalised variables} and the \textit{expansion normalised time} $\tau$ by demanding
\begin{align}
    \frac{1}{\si{d} \tau} = \frac{1}{D} \frac{1}{\si{d} t} \, ,
\end{align}
where $D$ is called \textit{dominant variable}. Usually $D$ is a function of $H$. Using the dominant variable one defines the expansion normalised variables by
\begin{align}
    H_D := \frac{H}{D} \, , \quad \sigma_D := \frac{\sigma}{D} \, , \quad \Omega_D:= \frac{\Omega}{D} \, .
\end{align}

\subsection{Energy conditions in General Relativity}\label{section_energy_conditions}
There exist several conditions for the energy momentum tensor, which are valid for different matter models. Some of them give necessary conditions for physical relevant matter models, others give useful additional assumptions, which are of importance for various conjectures. In this section we give a summary of the most important ones.
\begin{enumerate}
\item \textit{Weak energy condition}: It states, \[T_{\mu \nu} v^\mu v^\nu \ge 0\] for all timelike vectors $v$. In particular this means that the energy density of the matter is non-negative. For physically reasonable models this assumption is necessary.\cite[cf. page 218]{wald1984general} \label{weak_energy_condition}
\item \textit{Strong energy condition}: It states that for all timelike vectors $v$ we have \[T_{\mu \nu} v^\mu v^\nu \ge \frac{1}{2} \si{Tr}(T) g_{\mu \nu}v^\mu v^\nu \, .\] The formulation is equivalent to the condition that \[\rho + \si{tr}S \ge 0 \, .\] In particular, it means that the components of the principal pressures (the eigenvalues of $S_{ab}$) have a lower bound. In the mathematical literature this condition is also known as \textit{timelike convergence condition}.\cite[cf. page 219]{wald1984general} \label{strong_energy_condition}
\item \textit{Dominant energy condition}: It states that for all future directed timelike vectors $v$, the expression $-T\indices{^\mu_\nu} v^\nu$ is future directed timelike or null, cf. page 219 \cite{wald1984general}. It is equivalent to the inequality \[T(e_0,e_0) \ge |T(e_\alpha,e_\beta)| \, ,\] for any frame $\{e_\alpha\}$, cf. II.B \cite{heinkeli}. It is a condition which is believed to hold true for physical relevant matter models, as it implies that the velocities of matter are bounded by the speed of light.  The dominant energy condition implies the weak energy condition.\label{dominant_energy_condition}
\item \textit{Non-negative pressure condition}: This condition demands that the principal pressures are all non-negative. An alternative formulation is given by \[ 
T_{\mu \nu} z^\mu z^\nu \ge 0 \, ,
\] for all spacelike vectors $z$. From a physical point of view this means that pressures contribute more to the contraction of a spacetime rather than its expansion. For example Vlasov matter satisfies this condition as it models collisionless particles. Hence, if they are contracted they wont generate any forces acting against the contraction. A counter example is the energy momentum tensor of the electromagnetic field. In general it does not satisfy this condition.\cite[cf. II.B]{heinkeli} \label{non_negative_pressure}
\item \textit{Non-negative pressure-sum condition}: It is another less strict version of the non-negative pressure condition. It states that the sum of the principal pressures has to be non-negative. \label{non_negative_pressure_sum}
\end{enumerate}
\begin{figure}[th]
    \centering
    \includegraphics[width=0.5\textwidth]{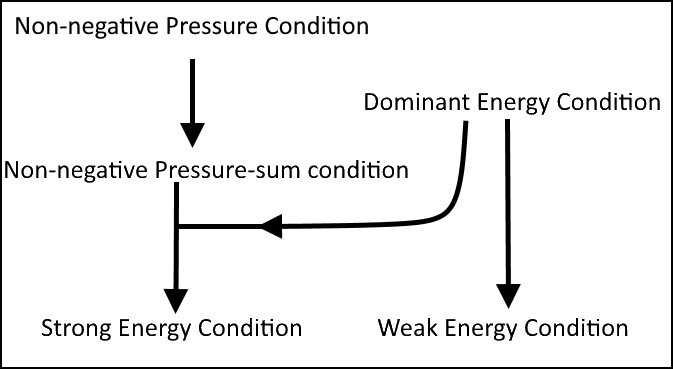}
    \caption{Diagram showing the hierarchy of the energy conditions.}
    \label{figure_energy_condition}
\end{figure}

Finally, we want to remark that the strong energy condition does not imply the weak energy condition, although the name may suggest it. The energy conditions \ref{weak_energy_condition}-\ref{dominant_energy_condition} are mostly independent from each other. Only the dominant energy conditions implies the weak energy condition. Regarding the pressure conditions: If the pressures are non-negative it follows trivially that also their sum is non-negative. The other way round is not true of course. If at least the non-negative pressure-sum condition is satisfied we have that $\si{tr}S \ge 0$ and hence, if the energy-density is positive the strong energy condition follows. In particular, the non-negative pressure-sum condition together with the dominant energy condition also implies the strong energy condition and the weak energy condition. Figure \ref{figure_energy_condition} shows the hierarchy of the energy conditions.

{\small
\bibliographystyle{alpha}
\bibliography{bibliography}}

\end{document}